\documentclass[twocolumn,aps,prl,superscriptaddress]{revtex4-2}
\usepackage{amsmath,empheq}
\usepackage{amssymb}
\usepackage{mathrsfs}
\usepackage{amsfonts}
\usepackage{booktabs}
\usepackage{physics}
\usepackage{graphicx}
\usepackage{calc}
\usepackage{epsfig}
\usepackage{color}
\usepackage{epstopdf}
\usepackage{xspace}
 \usepackage{rotating}
 \usepackage{appendix}
 \usepackage{chemformula}
 \usepackage{bm}

\usepackage[normalem]{ulem}

\usepackage[colorlinks=true,linkcolor=blue,citecolor=blue,filecolor=blue,urlcolor=blue]{hyperref}

\newcommand{\ttJJ}{$t$-$t'$-$J$-$J'$\xspace}

 \definecolor{bondiblue}{rgb}{0.0, 0.58, 0.71}
\begin{document}

\title{Coexistence of antiferromagnetism and topological $s+d+id$ superconductivity in twisted cuprate bilayers}
\author{Javier Castro Luaces}
\affiliation{Departamento de F\'isica Te\'orica de la Materia Condensada, Universidad Aut\'onoma de Madrid, E-28049 Madrid, Spain}
\affiliation{Condensed Matter Physics Center (IFIMAC),  Universidad Aut\'onoma de Madrid, E-28049 Madrid, Spain}
\affiliation{Instituto Nicol\'as Cabrera, Universidad Aut\'onoma de Madrid, E-28049 Madrid, Spain}
\author{Manuel Fern\'andez L\'opez}
\affiliation{Departamento de F\'isica Te\'orica de la Materia Condensada, Universidad Aut\'onoma de Madrid, E-28049 Madrid, Spain}
\affiliation{Condensed Matter Physics Center (IFIMAC),  Universidad Aut\'onoma de Madrid, E-28049 Madrid, Spain}
\affiliation{Instituto Nicol\'as Cabrera, Universidad Aut\'onoma de Madrid, E-28049 Madrid, Spain}
\affiliation{Institute of Physics, National Yang Ming Chiao Tung University, Hsinchu 30010, Taiwan}
\author{Jorge Bravo-Abad}
\affiliation{Departamento de F\'isica Te\'orica de la Materia Condensada, Universidad Aut\'onoma de Madrid, E-28049 Madrid, Spain}
\affiliation{Condensed Matter Physics Center (IFIMAC),  Universidad Aut\'onoma de Madrid, E-28049 Madrid, Spain}
\affiliation{Instituto Nicol\'as Cabrera, Universidad Aut\'onoma de Madrid, E-28049 Madrid, Spain}
\author{Jaime Merino}
\affiliation{Departamento de F\'isica Te\'orica de la Materia Condensada, Universidad Aut\'onoma de Madrid, E-28049 Madrid, Spain}
\affiliation{Condensed Matter Physics Center (IFIMAC),  Universidad Aut\'onoma de Madrid, E-28049 Madrid, Spain}
\affiliation{Instituto Nicol\'as Cabrera, Universidad Aut\'onoma de Madrid, E-28049 Madrid, Spain}
\begin{abstract}
We investigate the interplay between antiferromagnetism and topological superconductivity in twisted cuprate bilayers.
We demonstrate how topological $s+d+id$ superconductivity is induced by a twist of $\theta \sim 45^\circ$ in a microscopic $t$-$J$ type model of cuprate bilayers, where it robustly coexists with antiferromagnetic order at low hole doping.
This novel pairing state becomes physically accessible at realistic interlayer couplings in the overdoped regime. 
Crucially, it displays a characteristic 
$2$-fold rotational symmetry in angle-resolved photoemission spectroscopy (ARPES) not observed in the competing time-reversal symmetry breaking $d+id$ or $s+id$ states previously found. A finite Josephson effect at $45^\circ$ twist together with signatures in scanning tunneling spectroscopy of its two Majorana edge modes identifies this topological $s+d+id$ state unambiguously.
\end{abstract}
 \date{\today}
\maketitle

{\it Introduction.}---The intersection of unconventional superconductivity, magnetic order, and non-trivial topology in twisted two-dimensional materials reopens long-standing issues
regarding phase competition in quantum materials. Experimental advances in van der Waals (vdW) stacking have allowed the realization of twisted structures of graphene and transition metal dicalchogenides  
displaying unconventional moir\'e superconductivity~\cite{cao2018_a,cao2018_b,yu2019,xia2025} and interfacial high-T$_c$ superconductivity~\cite{kim_science_2023} in twisted cuprates~\cite{confalone2025}.
Josephson tunneling on twisted interfaces of stacked vdW $\text{Bi}_2\text{Sr}_2\text{CaCu}_2\text{O}_{8+x}$ (Bi-2212) crystals displays a strong suppression of the critical current around $45^\circ$, consistent with $d$-wave pairing~\cite{lee2021,kim_science_2023,vinokur2023,talantsev2024} and time-reversal symmetry breaking (TRSB)~\cite{kim_science_2023,basu2026}. However, competing experiments on twisted Bi-2212~\cite{zhu_prx_2021} and $\text{Bi}_2\text{Sr}_{2-x}\text{La}_x\text{CuO}_{6+y}$ (Bi-2201) ~\cite{wang_natcomm_2023} find a robust Josephson coupling near $45^\circ$. This provides a sharp contradiction that suggests a sizable isotropic $s$-wave component  in the interfacial superconductivity~\cite{zhu2023}, raising unsettled questions about electronic correlations and their interplay with other electronic orders~\cite{pixley2026} as antiferromagnetism (AFM).

Experiments have been mostly interpreted in terms of continuum models which assume $d$-wave superconductivity in each layer neglecting the strong electronic correlations and AFM expected in cuprates. Mean-field BCS-type approaches not only predict topological chiral $d+i d$ superconductivity~\cite{franz_natphys_2021,panda2026} around $\theta \sim 45^\circ$ twists but also, more recently, mixed $s/d$ topological and chiral superconductivity, $s+d_1 e^{i\phi_1} + d_2 e^{i \phi_2}$~\cite{li_arxiv_2026}. 
Approaches based on the $t$-$J$ model~\cite{ogata2008} capturing electronic correlation effects
find either non-topological chiral gapless superconductivity~\cite{ashvin2022} or a topological $d+id$ gapped superconductor~\cite{spalek_prb_2023} with no evidence of an
$s$-wave component around $45^\circ$ twist.
Numerical approaches on electron doped paramagnetic twisted Hubbard layers predict $d+id$ superconductivity at weak coupling~\cite{senechal2026,senechal2024,senechal2022}, {\it i.e.}, away from the regime relevant to hole-doped cuprates.
It is therefore crucial to determine and fully characterize the possible superconducting states emerging in twisted cuprate bilayers taking into account the strong electronic correlations, AFM, and spatial inhomogeneities occurring in cuprates, shedding light on the seemingly contradictory experimental observations. 
\begin{figure*}[t!]
\centering
\includegraphics[width=0.95\textwidth, clip]{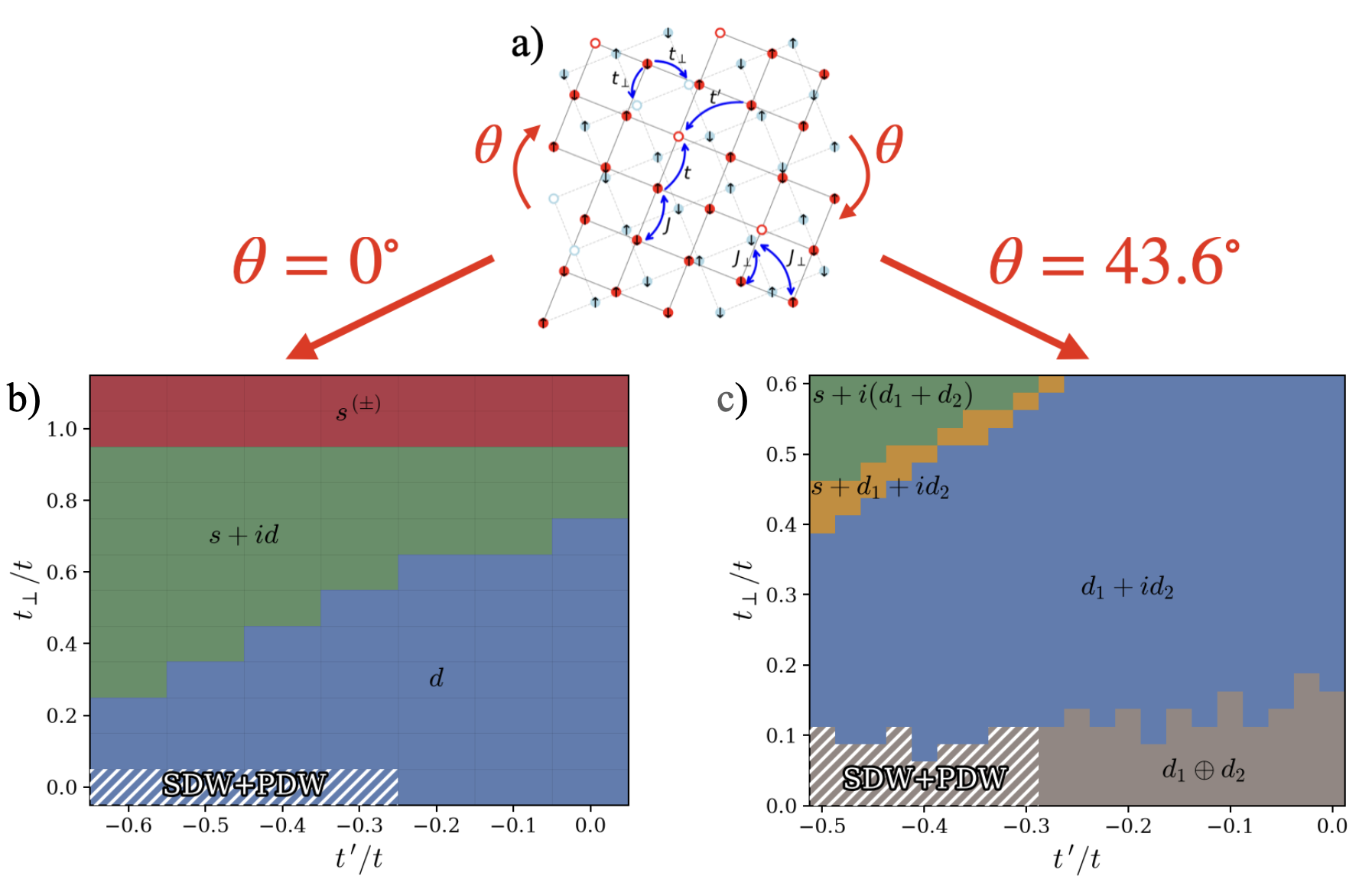}
\caption{Topological $s+d+id$ superconductivity in the $t$-$t'$-$t_\perp$-$J$-$J_\perp$ model for twisted bilayer cuprates at $\delta=1/8$. a) Sketch of the model indicating the intralayer ($t$, $t'$, $J$) and interlayer ($t_\perp$, $J_\perp$) couplings. b) The $t_\perp$ vs. $t'$ phase diagram of the untwisted bilayer, $\theta=0^\circ$. A non-topological TRSB $s+id$ superconductor emerges between $s^{(\pm)}$ and conventional $d$-wave superconductivity induced by $t_\perp$. The SDW+PDW phase stands for a combined spin density wave (SDW),  and a pair density wave (PDW). c) The $t_\perp$ vs. $t'$ phase diagram at $\theta=43.6^\circ$ twist. Topological $s+d_1+i d_2$ superconductivity emerges between $d_1+id_2$ and $s+i(d_1+d_2)$ phases. All superconducting phases coexist with AFM order. We fix $J=0.3t$.
}\label{fig:fig1}
\end{figure*}

In this Letter, we investigate the interplay between topological superconductivity, AFM and inhomogeneous electronic order within a realistic strongly correlated model for twisted cuprate bilayers. We employ an extended Gutzwiller wavefunction approximation (GWA), explicitly incorporating AFM, to solve a microscopic 
$t$-$t'$-$t_{\perp }$ -$J$-$J_\perp$ model of twisted bilayer cuprates effectively reaching the thermodynamic 
limit—a regime inaccessible to exact numerical approaches due to the large moir\'e unit cells. 
 At twist angles $\theta \sim 45^\circ$, we find an $s+d+id$ topological superconductor which arises as a compromise between the topological $d+id$ and the non-topological $s+id$ phases coexisting with AFM at weak hole doping, $\delta \lesssim 0.15-0.16$. The $s+d+id$ survives up to the overdoped regime at physically relevant interlayer couplings and 
 displays a characteristic lowering from $4$-fold to $2$-fold symmetry in ARPES experiments. In combination with the Josephson effect, and spatially resolved tunneling spectroscopy (STS) experiments, ---the latter probing its chiral Majorana edge modes---the $s+d+id$ state can be unambiguously distinguished from the previously discussed $s+id$ and $d+id$ states.

{\it Model and method.}---We consider a $t$-$t'$-$t_\perp$-$J$-$J_\perp$ model for twisted cuprate bilayers:
\begin{equation}
\mathcal{H} = 
 -\sum_{ij,\sigma} t_{ij}\, P_G \left( c_{i\sigma}^\dagger c_{j\sigma} + \text{h.c.} \right)P_G
+  \sum_{ij} J_{ij}\,  \mathbf{S}_{i} \cdot \mathbf{S}_{j},
\label{eq: main Hamiltonian}
\end{equation}
with: $P_G=\prod_i(1-n_{i\uparrow}n_{i\downarrow})$ the Gutzwiller projector \cite{supp}, the $t_{ij}=t (t'), \text{for } i,j  \in \text{n.n.(n.n.n.) sites in layer } l$, $t_{ij}=t^{(\perp)}_{ij},  i \in \text{ layer } l=1,j \in \text{ layer } l=2, $ with $t^{(\perp)}_{ij}=t_\perp\, e^{-(|\mathbf{r}_i-\mathbf{r}_j|-d)/r_0}$ decaying exponentially with the distance between sites on different layers, $d$ being the distance between layers and $r_0$ the characteristic interlayer decay length. The magnetic exchange couplings, $J_{ij}= {4 t_{ij}^2 \over U} $ with $U$ the Coulomb repulsion imply interlayer couplings $J^{(\perp)}_{ij} = (t_{ij}^{(\perp)}/t)^2 J$ with $J=4 t^2/U$, the n.n. AFM exchange. We assume $J'=0$.
The
variational wave function is constructed by projecting
out doubly occupied configurations from an uncorrelated
ansatz, $|\Psi_G \rangle  = P_G | \Psi_0 \rangle$. Its energy is evaluated within
an extended GWA that allows for AFM order~\cite{himeda_ogata2002,yang_zhang2009,tu2016}. The resulting
variational energy reads:
\begin{eqnarray}
&&E_{\text{var}}=\frac{\langle\Psi_G|\mathcal{H}|\Psi_G\rangle}{\langle\Psi_G|\Psi_G\rangle} = - \sum_{ij,\sigma} t_{ij}g_{ij}^{(t)} \langle c_{i\sigma}^\dagger c_{j\sigma} \rangle_0 +\text{c.c.}
\nonumber \\
&+&  \sum_{ij} J_{ij}\left( \tfrac{1}{2} g_{ij}^{(s,xy)} \langle S_i^+ S_j^- + S_i^- S_j^+ \rangle_0 + g_{ij}^{(s,z)} \langle S_i^z S_j^z \rangle_0 \right),
\nonumber \\
\label{eq:evar}
\end{eqnarray}
where the Gutzwiller renormalization factors, $g_{ij}$, capturing electronic correlation effects are functions of local occupations, $n_i$, magnetic moments, $m_i$, non-local bond, $\chi_{ij,\sigma}$, and symmetric and antisymmetric pairing amplitudes, $\Delta_{ij}^{(s)}$, $\Delta_{ij}^{(a)}$:
$n_{i\sigma} =\langle c_{i\sigma}^\dagger c_{i\sigma} \rangle_0$, $m_i = {1 \over 2} \langle (n_{i\uparrow}-n_{i\downarrow}) \rangle_0 $, $ \Delta^{(s)}_{ij} =  \frac{1}{2} \langle \left( c_{i\uparrow}^\dagger c_{j\downarrow}^\dagger - c_{i\downarrow}^\dagger c_{j\uparrow}^\dagger \right) \rangle_0 $, $ \Delta^{(a)}_{ij} =  \frac{1}{2} \langle \left( c_{i\uparrow}^\dagger c_{j\downarrow}^\dagger + c_{i\downarrow}^\dagger c_{j\uparrow}^\dagger \right) \rangle_0 $, $ \chi_{ij,\sigma} = \langle c_{i\sigma}^\dagger c_{j\sigma} \rangle_0$.  Minimization of the variational energy $E_{\text{var}}+\mu \sum_i n_i$ (with $\mu$ the chemical potential fixing $\delta$) w.r.t.  \{ $\chi_{ij,\sigma}$, $\Delta^{(s)}_{ij}$, $\Delta^{(a)}_{ij}$, $n_{i\sigma}$\} is equivalent to finding the ground state, $|\Psi_0 \rangle$, of the effective Hamiltonian~\cite{yang_zhang2009,fawang2024}:
\begin{equation}
\mathcal{H}_{\text{eff}} = \sum_{ij, \sigma} \frac{\partial W}{\partial \chi_{ij,\sigma}} \hat{\chi}_{ij,\sigma} 
+ \sum_{\substack{ij \\ \alpha=s,a}}\frac{\partial W}{\partial \Delta_{ij}^{(\alpha)}} \widehat{\Delta}_{ij}^{(\alpha)}+\text{h.c.}
\nonumber \\
+ \sum_{i, \sigma}  \frac{\partial W}  {\partial n_{i\sigma}}\hat{n}_{i\sigma},
\nonumber \\
\end{equation}
where $W=E_\text{var}+\mu\sum_i n_i$, with $\mu$ the chemical potential.
The resulting self-consistent equations are solved at $T=0$ on moiré unit cells with periodic boundary conditions (PBC) reaching the thermodynamic limit.~\cite{supp} 
\begin{figure}[htbp]
    \centering
    \hspace{-1cm}
\includegraphics[width=0.45\textwidth,clip]{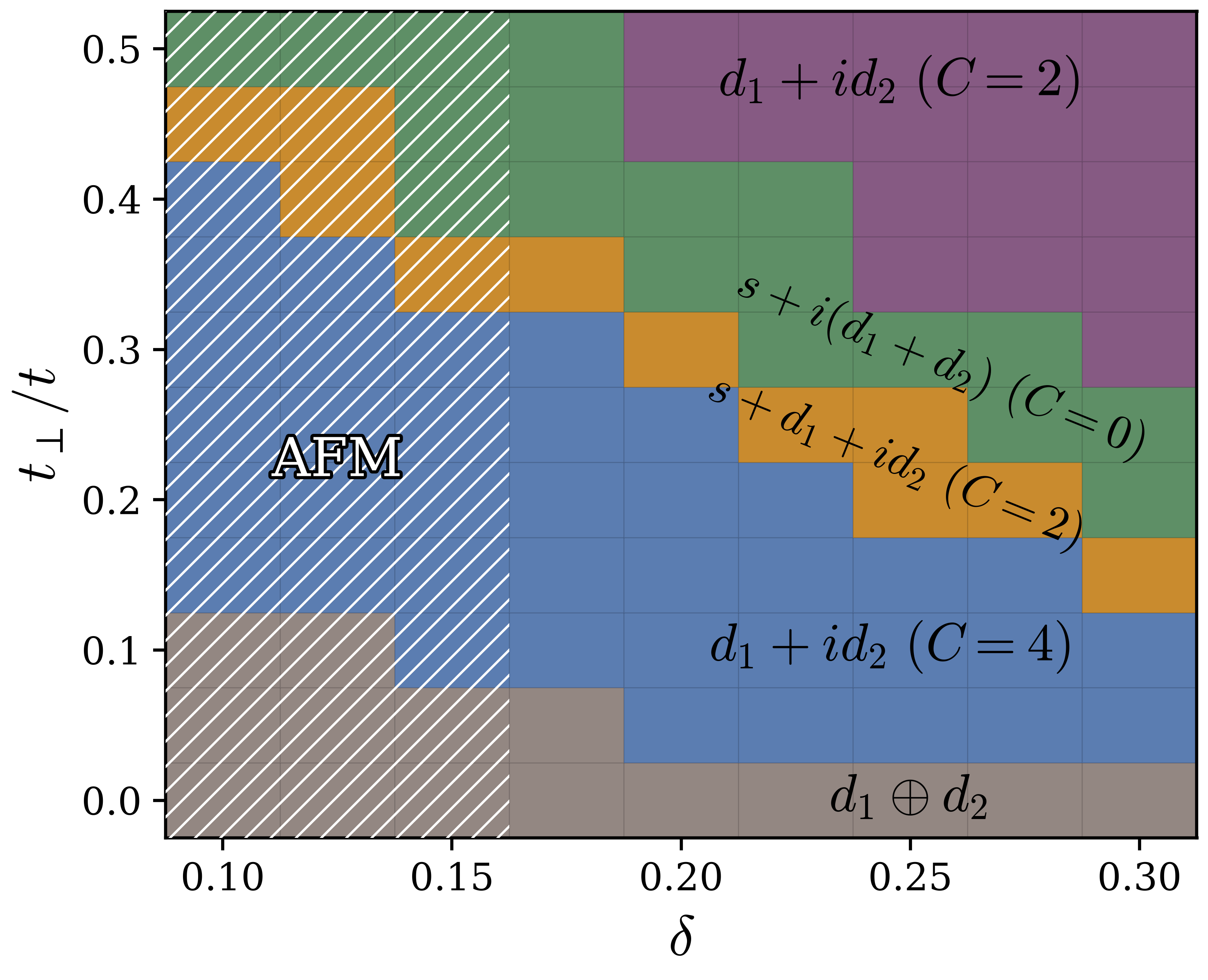}

\caption{Doping dependence of topological superconductivity in the twisted cuprate bilayer at twist angle, $\theta=43.6^\circ$. The $t_\perp$ vs. $\delta$ phase diagram shows the various spatially uniform superconducting phases and the region (hatched) in which they coexist with AFM order. $d_1\oplus d_2$ denotes effectively decoupled $d$-wave layers with no phase locking between them, and $C$ the Chern number of the superconducting phases with broken time-reversal symmetry. Note how the $s+d_1+id_2$ phase is stabilized at progressively lower $t_\perp$ as hole doping, $\delta$, increases. We fix $J=0.3t$ and $t'=-0.5t$ in this plot.
}
\label{fig:fig2}
\end{figure}

\textit{Phase diagram of bilayer cuprates at $\delta = 1/8$: effect of twist.}---We analyze the twist dependence of the $t$-$t'$-$t_\perp$-$J$-$J_\perp$ model at $\delta = 1/8$ hole doping, first considering uncoupled ($t_\perp/t = 0$), untwisted layers [Fig.~1(b)]. We find robust $d_{x^2-y^2}$-wave superconductivity for $-0.25 < t'/t \le 0$ and a SDW+PDW phase for $t' < -0.25t$. This is consistent with recent 8-leg cylinder DMRG calculations on the $t$-$J$ model at $\delta \sim 0.1-0.2$~\cite{sheng_2025}, but contrasts with Hubbard model DMRG studies~\cite{kivelson_deveraux_2026} indicating superconductivity suppression by CDW/SDW order. Such discrepancies stem from finite size and boundary effects in DMRG, whereas our GWA approach effectively reaches the thermodynamic limit. Indeed, our results are consistent with recent neural-network quantum states calculations~\cite{georges_2026} which find robust $d$-wave superconductivity at $\delta=1/8$, $t'=-0.2t$, and $U=8t$, as well as with previous combined DMRG+AFQMC calculations showing a coexistence of superconductivity with partially filled stripes~\cite{zhang_white_2024}. Additionally, we find a PDW~\cite{fradkin_2015,agterberg_2020} for $t' < -0.25t$, characterized by a weak spatial modulation of the pairing amplitude intertwined with CDW and SDW orders coexisting with uniform $d$-wave superconductivity, {\it i.e.} a SDW+PDW reminiscent of the PDW recently detected in 214 cuprates near $1/8$ doping~\cite{chen_2025}. Weak (robust) PDWs arise in 8-leg cylinder DMRG~\cite{sheng_2025} (infinite tensor networks~\cite{zheng_2025}) calculations of the \ttJJ model consistent with previous VMC calculations~\cite{himeda_ogata2002} and with the Gutzwiller renormalized mean-field approach including AFM~\cite{ogata_himeda2003,yang_zhang2009} used here. 
Our thermodynamic limit results of Fig.~\ref{fig:fig1}b), consistent with genuinely 2D infinite methods show that the PDW is favored by the flat band sections arising at $t'=-0.5t$. On the other hand, there is yet no evidence of PDWs in the Hubbard model from DMRG~\cite{kivelson_deveraux_2026}. Increasing $t_\perp/t$ in untwisted layers (see Fig.~\ref{fig:fig1}b)) leads to non-topological TRSB $s+id$ superconductivity finally becoming a $s^{(\pm)}$-wave superconductor at large $t_\perp \gtrsim 0.9t$ independent of $t'/t$, consistent with previous works~\cite{maier_scalapino_bilayers_2011, weiwu_high-tc_2024}. On the other hand, the SDW+PDW state survives to very small $t_\perp <0.075 t$.  

Twisting the layers at $\theta = 43.6^\circ$ leads to 
various TRSB pairing states including $d+id$ topological superconductivity. Here and below, $d_l$ denotes the $d_{x^2-y^2}$ pairing amplitude of layer $l=1,2$, each defined in its own crystallographic frame (so that for $\theta\simeq45^\circ$, $d_2$ acts as a $d_{xy}$-like component in the frame of layer 1), and $d_1+id_2$ stands for a relative phase $\phi_2-\phi_1=\pm\pi/2$ between the two layers. The $d_1+id_2$ phase at hole doping $\delta=1/8$, occupies a broad region of the phase diagram (see Fig.~\ref{fig:fig1}c)) consistent with pioneering mean-field BCS-type work~\cite{franz_natphys_2021,pixley2026} and numerical work on $t$-$J$ type models~\cite{spalek_prb_2023}.
Most importantly, electronic correlations also give rise to a $s + d_1 e^{i \phi_1} + d_2 e^{i \phi_2} $ topological superconductor, with $\phi_1=0$ and $\phi_2 \approx \pi/2$. This state consists of an $s_1+d_1$ contribution in layer 1 and a $s_2+id_2$ in layer 2 with $s_1=s_2$ breaking the combined $C_4 U(1)$ symmetry on both layers consistent with macroscopic Ginzburg-Landau arguments at dominant biquadratic $d-d$ couplings~\cite{li_arxiv_2026}: $g_{sd}<g_{dd}$. The $s+d_1+id_2$ is favored by sufficiently large $t_\perp \gtrsim  0.4t$ and $t' \lesssim -0.25 t$.

{\it Effect of doping on AFM and topological superconductivity in twisted layers.}--- AFM and $d$-wave superconductivity coexist in the $t$-$J$ model of a single CuO$_2$ layer of the cuprates. The GWA method extended to include AFM~\cite{himeda_ogata1999} predicts this coexistence for $\delta \lesssim 10 \% $, consistent with VMC calculations~\cite{ogata_himeda2003}. Similarly, coexistence between AFM and topological superconductivity is found, and is robust against $t_\perp$, as shown in Fig.~\ref{fig:fig2} for
$\theta=43.6^\circ$. Such coexistence persists up to a critical doping, $\delta_c \approx0.16-0.17$, at which the AFM moment, $m$, is completely suppressed. The gradual suppression of $m$ with $\delta$~\cite{supp} indicates a  microscopic and homogeneous coexistence between AFM and topological superconductivity for $\delta < \delta_c$ and a pure topological superconductor for $\delta> \delta_c$. Note also that $m$ is slightly enhanced by $t_\perp$~\cite{supp}, presumably reflecting the interlayer exchange $J_\perp\propto t_\perp^2$, which reinforces N\'eel correlations between the layers. Concomitantly, the hopping renormalization factors,
{$g_{ij}^{(t)}$,} increase monotonically with doping displaying a kink at 
$\delta_c$~\cite{supp}.
\begin{figure}[htbp] 
    \centering
    \includegraphics[width=0.5\textwidth]{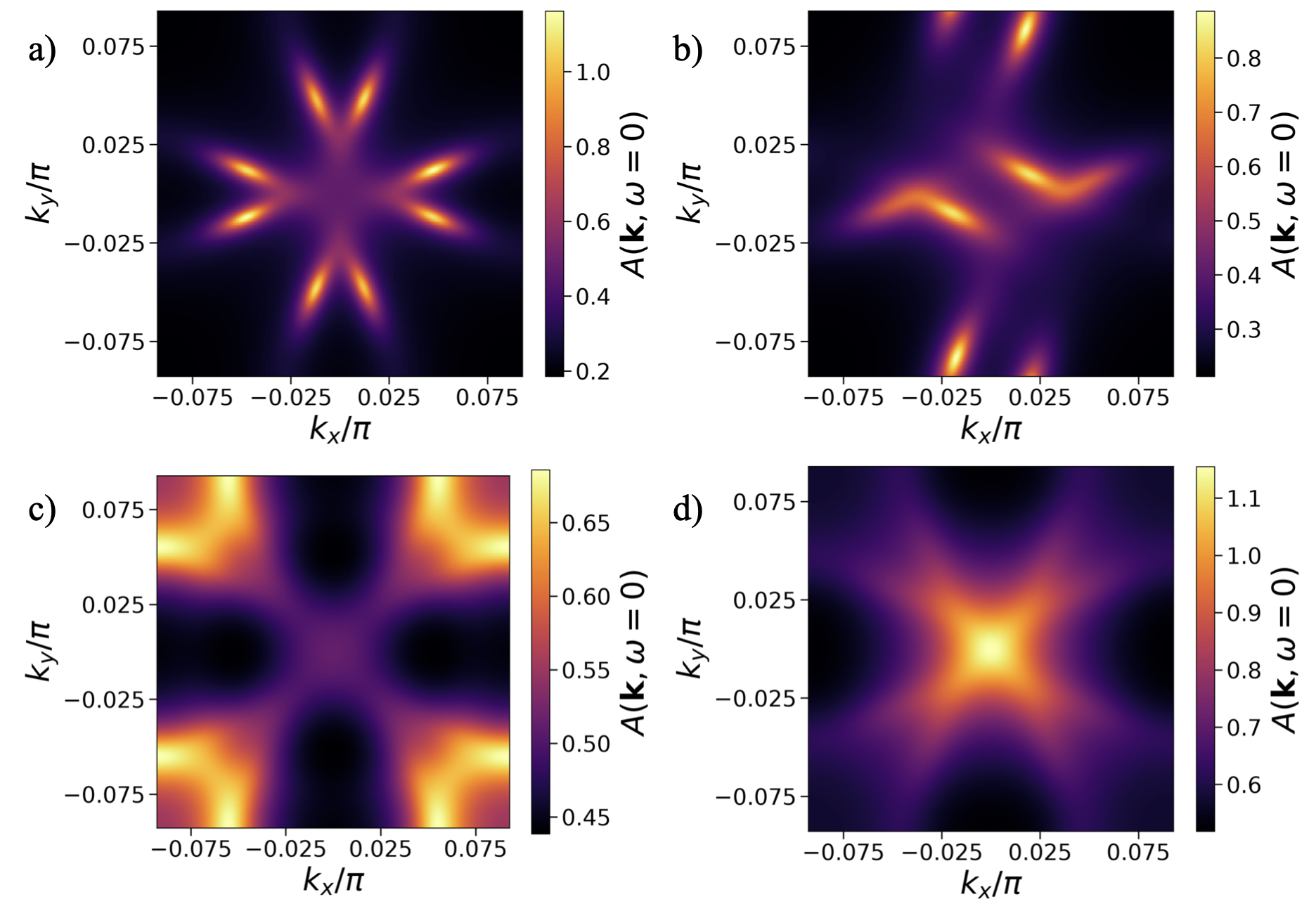} 
    \caption{Characteristic $2$-fold rotational symmetry of the $s+d+id$ pairing state in ARPES. The spectral function at the Fermi level, $A(\mathbf{k},\omega=0)$, in the Moir\'e Brillouin zone is displayed for (a) $d_1+id_2$ ($C=4$), (b) $s+d_1+id_2$ ($C=2$), (c) $s+id$ ($C=0$) and (d) $d_1+id_2$ ($C=2$) pairing states. Crucially, the $s+d_1+id_2$ state is the only one whose spectral function $A(\mathbf{k},0)$ breaks $C_4$ rotational symmetry. The parameters used are: $t'=-0.5t$, $J=0.3t$, with (a) $\delta=1/8$, $t_\perp=0.35t$, (b) $\delta=1/8$, $t_\perp=0.42t$, (c) $\delta=1/5$, $t_\perp=0.4t$, and
    (d) $\delta=1/5$, $t_\perp=0.45t$.}
    \label{fig:fig3}
\end{figure}

As hole doping increases, the topological superconducting phases ($s+d_1+id_2$, $d_1+id_2$) are stabilized at progressively lower $t_\perp$, as shown in Fig.~\ref{fig:fig2}. In the single-layer $t$-$J$ model, it is well known that $d$-wave superconductivity is suppressed with doping above $\delta\sim 0.1-0.15$, disappearing completely around $\delta \sim 0.3$.~\cite{himeda_ogata2002}
This is because the renormalized kinetic energy, $g^{(t)}t$, grows with $\delta$ while the exchange scale remains fixed, so that pairing becomes progressively less favorable and the system evolves towards a weakly correlated metal.
From our results in Fig.~\ref{fig:fig2}, at a fixed $t_\perp$ (e.g.\ $t_\perp=0.3t$), one can tune the system through the $d_1+i d_2$, $s+ d_1+i d_2$ and $s+i (d_1+d_2)$ states by increasing $\delta$, until a metallic phase is encountered (not shown). At sufficiently large $t_\perp$ and $\delta$, the $C=4$, $d_1+id_2$ state featuring eight gapped Dirac cones, becomes a $C=2$, $d_1+id_2$ due to bonding-antibonding splitting effectively leading to single layer $d_1+id_2$ superconductivity.~\cite{franz_natphys_2021,supp}

The $s+d_1+id_2$ phase occupies a narrow strip between the $d_1+id_2$ ($C=4$) and the topologically trivial $s+i(d_1+d_2)$ states and can be viewed as a compromise between them: it retains the $\pm\pi/2$ phase locking between the $d$-wave components of the two layers, and hence their chirality, while admitting a subdominant $s$-wave admixture. The change in topology between the two states requires the bulk gap to close, leading to this intermediate $C=2$ phase.

\begin{figure}[htbp]
\centering
\includegraphics[width=0.45\textwidth]{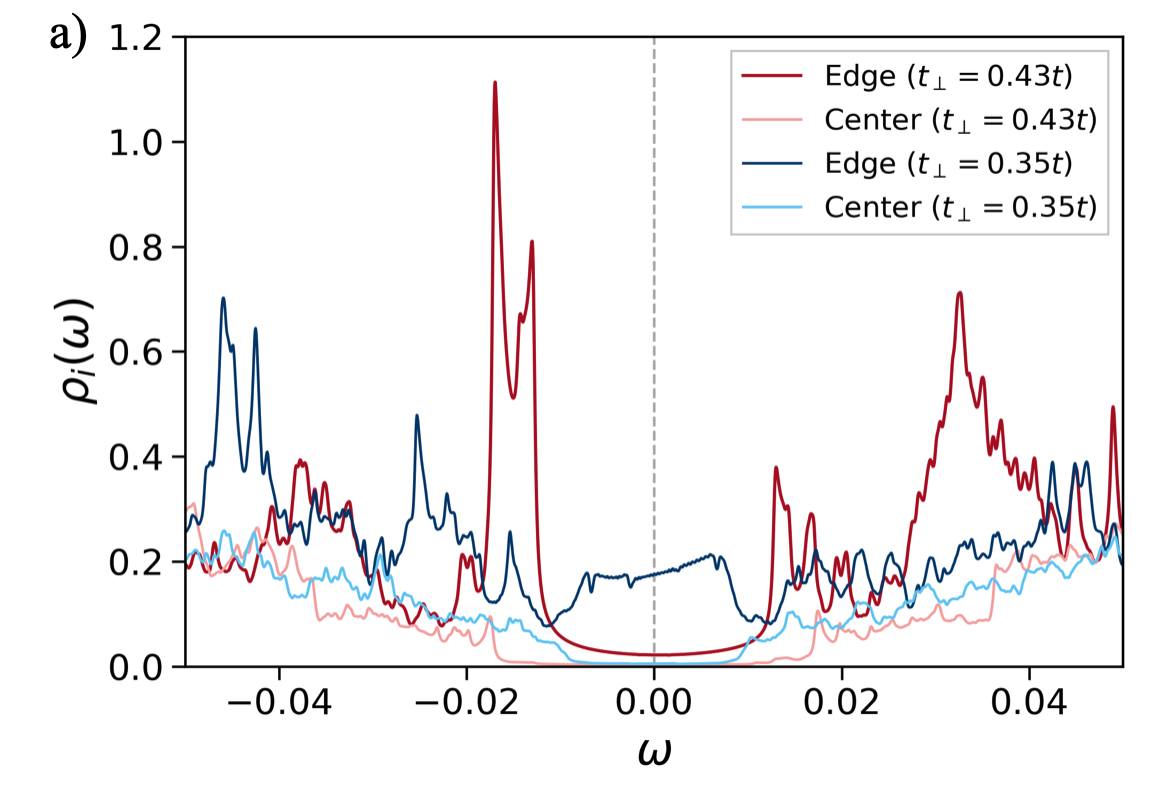}
\includegraphics[width=0.45\textwidth]{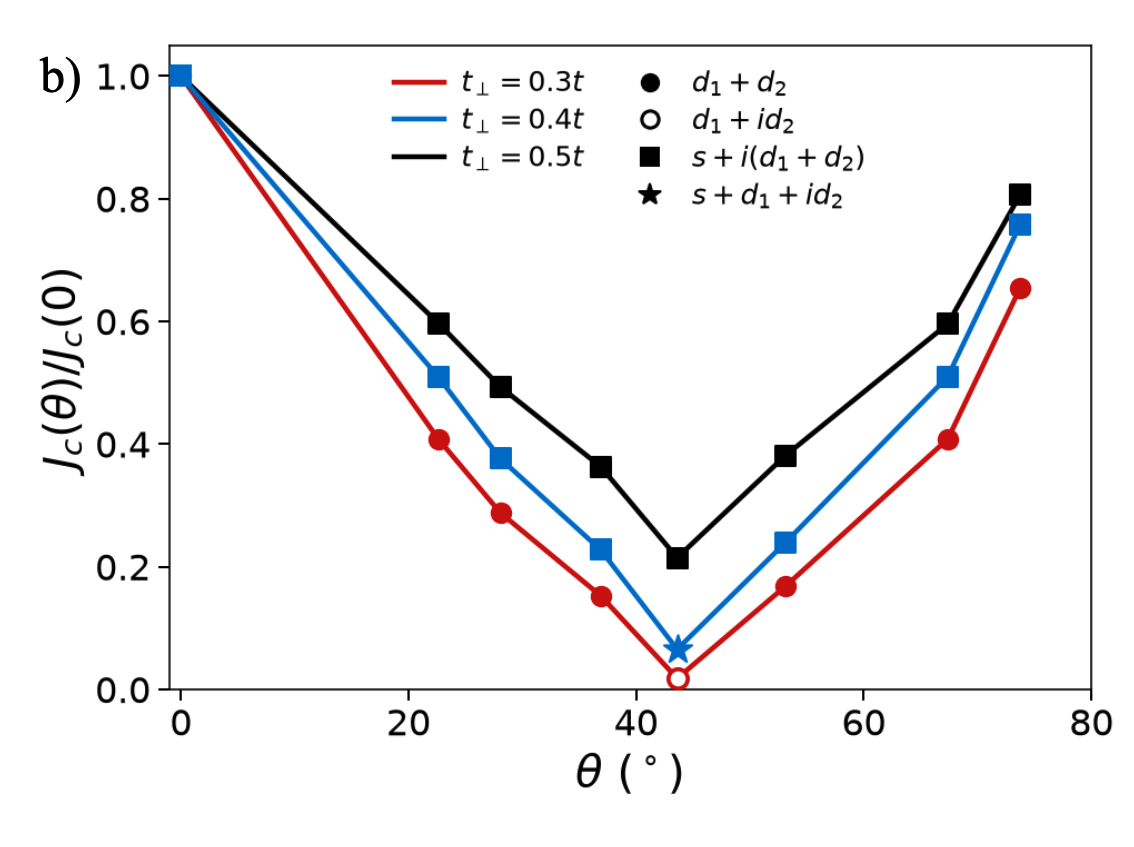}
\caption{Experimental signatures of $s+d+id$ superconductivity in twisted bilayers. (a) LDOS relevant to STS experiments at the edge (dark lines) and at the center (light lines) of a ribbon for the $C=4$, $d_1+id_2$ ($t_\perp=0.35t$, blue) and $C=2$, $s+d_1+id_2$ ($t_\perp=0.43t$, red) states at twist angle, $\theta =43.6^\circ$ are shown. 
(b) Twist angle dependence of the critical Josephson current, $J_c(\theta)$, normalized to its maximum value, for three different interlayer hopping amplitudes, $t_\perp$, with symbols identifying the superconducting state found at each angle. We have used, $\delta=1/8$, $t'=-0.5t$ and $J=0.3t$ in this plot.
}
    \label{fig:fig4}
\end{figure}

{\it  Experimental signatures of interfacial $s+d_1+i d_2$ superconductivity.}---To distinguish the emergent superconducting phases, we analyze their distinct momentum- and real-space structures detectable via ARPES and STS. The 4-fold rotational symmetry of the $d_1+id_2$ states is lowered to a $2$-fold symmetry in the $s+d_1+id_2$ state, owing to the constructive/destructive interference between the in-phase $s$- and $d$-wave components along the two antinodal directions. Such symmetry lowering is evident when comparing the zero-energy spectral functions at the Fermi surface, $A({\bf k}, 0)$, of the $d_1+id_2$ and $s+ d_1+i d_2$ states shown in Fig.~\ref{fig:fig3}a),b). Although these states are fully gapped, $A(\mathbf{k},0)$ is nonzero due to the finite broadening $\eta$ leading to finite spectra around the Fermi energy. Hence, the nematic character of the $s+ d_1+i d_2$ state should be directly visible in ARPES experiments~\cite{li_arxiv_2026}. 
Interestingly, the $s+i(d_1+d_2)$ state also displays fourfold rotational symmetry in $A(\mathbf{k},0)$, although its gap function is invariant only under twofold rotations. This follows from the invariance of the state under $C_4\mathcal{T}$, which also forces its Chern number to vanish~\cite{supp}.

The topological character can be probed through STS by laterally displacing the STM tip from the interior of the sample to its edges. In the bulk, all TRSB pairing states are fully gapped and hence qualitatively indistinguishable, as shown in Fig.~\ref{fig:fig4}a). However, the LDOS is enhanced as the tip approaches the sample edges of the topological pairing states, due to the associated Majorana chiral edge states. This enhancement is much larger for $d_1+id_2$ than $s+d_1+i d_2$ pairing (see Fig.~\ref{fig:fig4}a)). This is consistent with the bulk-boundary correspondence: the $C=4$, $d_1+id_2$ state hosts four chiral Majorana modes per edge, whereas the $C=2$, $s+d_1+id_2$ state hosts only two. In addition, the edge modes of the $d_1+id_2$ state are much less dispersive than those of the $s+d_1+id_2$ state~\cite{supp}, which further enhances the zero-bias LDOS in the former. Hence, a large zero-bias LDOS at the sample edges would point to the $d_1+id_2$ state, compared to the weaker, but finite, LDOS edge signal of the $s+d_1+id_2$ state. In spite of the suppressed zero-bias LDOS of the $s+d_1+id_2$ compared to the $d_1+id_2$ state, large side bands associated with its edge states emerge.   

An unambiguous state identification can be achieved by combining ARPES, STS, and the interlayer Josephson current. Fig.~\ref{fig:fig4}b) compares the twist-angle dependence of the critical current $J_c(\theta)$ for three $t_\perp$ values. 
For $t_\perp=0.3t$, both layers retain $d$-wave pairing at all twist angles, with a time-reversal-symmetric relative phase away from $45^\circ$, whereas near $45^\circ$, $|\phi_1-\phi_2|\simeq\pi/2$, giving rise to the $d_1+id_2$ state. Although $J_c(45^\circ) = 0$ for direct Cooper-pair tunneling between weakly coupled $d$-wave superconductors, the general current-phase relation including co-tunneling~\cite{tumuru2022,franz_natphys_2021}, $J_c(\theta \sim 45^\circ) = \max_{\phi}[J_{c1}(0)\cos(2\theta)\sin(\varphi) - J_{c2}\sin(2\varphi)]$, predicts a finite $J_c(45^\circ)$ for $J_{c2} > 0$, as found here. This second harmonic stabilizes the TRSB $d_1 + id_2$ ground state via double minima at $\varphi = \pm\pi/2$ in the free energy. At $t_\perp = 0.5t$, $J_c$ depends only weakly on $\theta$ due to the isotropic $s$-wave component of the $s+i(d_1+d_2)$ state. Crucially, at $t_\perp = 0.4t$, the system enters the $s + d_1 + id_2$ topological phase for $\theta = 43.6^\circ$ twist [star in Fig.~4b)], exhibiting a small but finite Josephson effect that lies between the $d_1 + id_2$ and the non-topological $s+i(d_1+d_2)$ states.

{\it Conclusions and outlook.}---We have found that $s+d_1+id_2$ interfacial topological superconductivity emerges in a realistic microscopic model of twisted cuprates and coexists at low hole doping with, rather than being suppressed by, antiferromagnetism. 
The $s+d_1+id_2$ state becomes within reach of realistic interlayer couplings in the overdoped regime. 
The characteristic nematic 2-fold symmetry in ARPES and the two Majorana edge modes detectable in STS allow its unambiguous identification. The agreement between our Gutzwiller approach and 2D tensor networks for the single-layer $t$-$J$ model~\cite{troyer_2014,zheng_2025} highlights the robustness of our predictions. 
Our results naturally reconcile seemingly contradictory twisted-junction experiments: depending on effective interlayer coupling and doping, the bilayer realizes either a chiral $d_1+id_2$ state with suppressed Josephson coupling near $45^\circ$~\cite{kim_science_2023}, or states with a sizable $s$-wave admixture and weakly $\theta$-dependent $J_c$~\cite{zhu_prx_2021,zhu2023,wang_natcomm_2023}.



\acknowledgments
We acknowledge financial support from (Grant No.
PID2022-139995NB-I00) MICIN/FEDER, Unión Europea,
from grant "Excelencia María de Maeztu" CEX2023-001316-M funded by MICIU/AEI /10.13039/501100011033).
J.C.L. acknowledges financial support from the FPI Grant No. PREX2023-000114. M.F.L. acknowledges financial support from Taiwanese NSTC Grant No. 115-2811-M-A49-549.
\nocite{*}
\bibliographystyle{apsrev4-1}
\bibliography{main}
\end{document}


\title{Supplementary material for Coexistence of antiferromagnetism and topological $s+d+id$ superconductivity in twisted cuprate bilayers }
\author{Javier Castro Luaces}
\affiliation{Departamento de F\'isica Te\'orica de la Materia Condensada, Universidad Aut\'onoma de Madrid, E-28049 Madrid, Spain}
\affiliation{Condensed Matter Physics Center (IFIMAC),  Universidad Aut\'onoma de Madrid, E-28049 Madrid, Spain}
\affiliation{Instituto Nicol\'as Cabrera, Universidad Aut\'onoma de Madrid, E-28049 Madrid, Spain}
\author{Manuel Fern\'andez L\'opez}
\affiliation{Departamento de F\'isica Te\'orica de la Materia Condensada, Universidad Aut\'onoma de Madrid, E-28049 Madrid, Spain}
\affiliation{Condensed Matter Physics Center (IFIMAC),  Universidad Aut\'onoma de Madrid, E-28049 Madrid, Spain}
\affiliation{Instituto Nicol\'as Cabrera, Universidad Aut\'onoma de Madrid, E-28049 Madrid, Spain}
\affiliation{Institute of Physics, National Yang Ming Chiao Tung University, Hsinchu 30010, Taiwan}
\author{Jorge Bravo-Abad}
\affiliation{Departamento de F\'isica Te\'orica de la Materia Condensada, Universidad Aut\'onoma de Madrid, E-28049 Madrid, Spain}
\affiliation{Condensed Matter Physics Center (IFIMAC),  Universidad Aut\'onoma de Madrid, E-28049 Madrid, Spain}
\affiliation{Instituto Nicol\'as Cabrera, Universidad Aut\'onoma de Madrid, E-28049 Madrid, Spain}
\author{Jaime Merino}
\affiliation{Departamento de F\'isica Te\'orica de la Materia Condensada, Universidad Aut\'onoma de Madrid, E-28049 Madrid, Spain}
\affiliation{Condensed Matter Physics Center (IFIMAC),  Universidad Aut\'onoma de Madrid, E-28049 Madrid, Spain}
\affiliation{Instituto Nicol\'as Cabrera, Universidad Aut\'onoma de Madrid, E-28049 Madrid, Spain}
\date{\today}
\begin{abstract}
\end{abstract}

\maketitle

\section{Model for twisted cuprate bilayers}

The low-energy physics of cuprates is predominantly governed by the CuO$_2$ planes. In the strongly correlated regime relevant to these materials, the Cu-O electronic degrees of freedom can be mapped onto an effective $t-J$ model on the Cu square lattice with lattice constant $a_0$, in which doped holes propagate through an antiferromagnetic spin background. We include a next-nearest-neighbor hopping amplitude $t'$ as a leading longer-range kinetic contribution. Since $t'$ connects sites within the same antiferromagnetic sublattice, it can substantially influence the competition between ordered phases. The resulting single-layer hamiltonian reads:
\begin{align}
\mathcal{H}_{\text{$t$-$t'$-$J$}} = &-t\sum_{\langle i, j\rangle,\sigma} P_G(c_{i\sigma}^\dagger c_{j\sigma}
+\text{h.c.})P_G\nonumber\\ &-t'\sum_{\langle\langle i, j\rangle\rangle,\sigma}P_G(c_{i\sigma}^\dagger c_{j\sigma}+\text{h.c.})P_G\nonumber\\
&+ J\sum_{\langle i, j\rangle} \mathbf{S}_i\cdot \mathbf{S}_j,
\label{eq: monolayer hamiltonian}
\end{align}
where $J=4t^2/U$ denotes the nearest-neighbor superexchange interaction, with $U$ being the strength of the on-site Coulomb repulsion. The next-nearest-neighbor exchange coupling $J'$ is assumed to be zero and neglected. $P_G=\prod_i(1-n_{i\up}n_{i\down})$ denotes the Gutzwiller projector.   
\begin{figure}[t!]
\centering
\includegraphics[width=10cm]{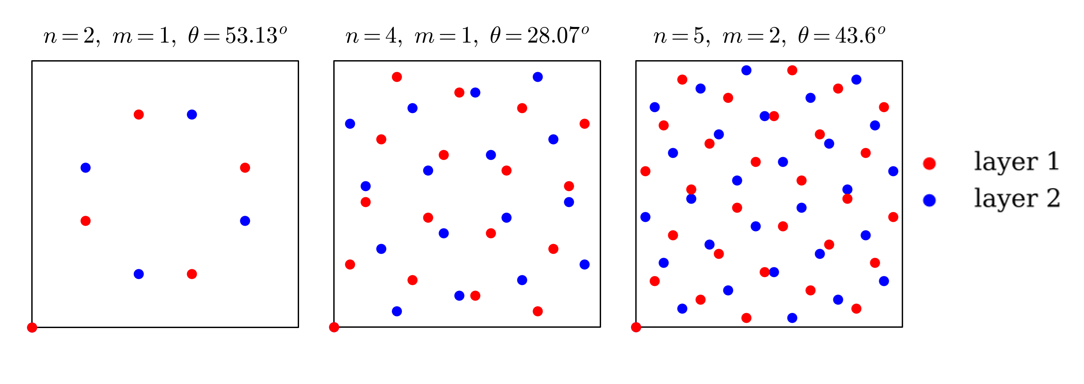}
\caption{Representative commensurate moiré unit cells for the twist geometries $(n,m)=(2,1),\ (4,1), \text{and} \ (5,2)$, shown from left to right. The square outlines delimit the corresponding primitive morié cells, while the red and blue dots denote the Cu sites in layers 1 and 2, respectively. Note that, to accommodate Néel order under periodic boundary conditions while preserving the moiré geometry, each primitive cell must be enlarged to form a $2\times2$ supercell.}
\label{fig: Twist lattices}
\end{figure}
To model the cuprate bilayer, we consider two stacked CuO$_2$ planes rotated by a relative angle $\theta$, each governed by Eq.~\ref{eq: monolayer hamiltonian}. To describe both aligned and twisted geometries within a unified framework, we introduce distance-dependent interlayer hopping and exchange interactions. The resulting bilayer Hamiltonian reads:
\begin{align}
    \mathcal{H}=&\sum_{l=1, 2}\mathcal{H}_\text{$t$-$t'$-$J$}^{(l)} - \sum_{ij,\sigma} t_{ij}^{(\perp)}P_G\left( c_{i1\sigma}^\dagger c_{j2\sigma} + \text{h.c.} \right)P_G 
    \nonumber \\
    &+ \sum_{ij}J_{ij}^{(\perp)}\mathbf{S}_{i1}\cdot\mathbf{S}_{j2},
\end{align}
where $l=1,2$ denotes the layer index.
Since the relative rotation modifies the local stacking configuration, we model the interlayer hopping as an isotropic, exponentially decaying function of the distance between sites in opposite layers:
\begin{align}
    t_{ij}^{(\perp)}=t_\perp\exp \left( -\frac{||\mathbf{r}_{i1}-\mathbf{r}_{j2}||-d}{r_0} \right),
\end{align}
where $\mathbf{r}_{il}$ denotes the three-dimensional position of site $i$ in layer $l$, $d=2.1a_0$ is the interlayer separation and $r_0=0.5a_0$ is the characteristic decay length. We restrict the interlayer couplings to site pairs separated by a distance smaller than a cutoff radius $r_c=3.1a_0$. For vertically aligned sites, $||\mathbf{r}_{i1}-\mathbf{r}_{j2}||=d$, yielding $t_{ij}^{(\perp)}=t_\perp$. The corresponding interlayer exchange interaction is determined by the strong-coupling relation:
\begin{align}
    J_{ij}^{(\perp)}=J\left( \frac{t_{ij}^{(\perp)}}{t} \right)^2.
\end{align}
By rewriting the bilayer Hamiltonian in the compact form:
\begin{equation}
    \mathcal{H}=-\sum_{ij, \sigma}t_{ij}P_G(c_{i\sigma}^\dagger c_{j\sigma} +\text{h.c.})P_G + \sum_{ij} J_{ij}\mathbf{S}_{i}\cdot \mathbf{S}_j,
\end{equation}
where the layer indices are absorbed into the site labels and the hopping and exchange couplings are defined as:
\begin{align}
&(t_{ij}, J_{ij}) = 
\begin{cases}
(t,4t^2/U), & i,j \in \text{n.n. in layer } l \\
(t',0), & i,j \in \text{n.n.n. in layer } l \\
(t^{(\perp)}_{ij}, 4t_{ij}^{(\perp)2}/U), & i \in   l=1, \, j \in l=2
\end{cases}
\nonumber
\end{align}
the expression for $\mathcal{H}$ provided in the main text is recovered.  

For the underlying square lattice, periodic moiré supercells arise at commensurate twist angles, characterized by two integers $n$ and $m$ through $\theta=2\arctan(m/n)$. The corresponding square supercell has a linear size $a=a_0\sqrt{n^2+m^2}$ and contains $n^2+m^2$ sites per layer, giving a total of $2(n^2+m^2)$ sites in the bilayer. 
For the numerical calculations, we consider systems comprising $N_x\times N_y$ moiré supercells with periodic boundary conditions (PBC), containing a total of $2(n^2+m^2)N_xN_y$ sites. The largest systems correspond to $N_x=N_y=51$ for the commensurate geometry $(n,m)=(10,4)$ ($\theta=43.6^\circ$). These system sizes are sufficiently large to describe the thermodynamic limit. 
Within this construction, the aligned bilayer is recovered for $n=m$, corresponding to a 90$^\circ$ rotation that leaves the underlying square lattice invariant. We use $(n,m)=(8,8)$ to represent this limit throughout our work.

\section{Gutzwiller Renormalized Mean-Field Theory (GRMFT)}

To determine the ground state of the system, we employ a Gutzwiller renormalized mean-field theory (GRMFT) extended to include magnetic order. Within this framework, strongly local correlations are incorporated by projecting an uncorrelated state $\ket{\Psi_0}$ onto the subspace without double occupancy:
\begin{align}
    \ket{\Psi_G} = \mathcal{P}_G\ket{\Psi_0}=\prod_i (1-n_{i\uparrow}n_{i\downarrow})\ket{\Psi_0},
\end{align}
where $\mathcal{P}_G$ denotes the Gutzwiller projector.

For the $t$-$J$ Hamiltonian, the variational energy takes the form:
\begin{eqnarray}
&&E_\text{var}=\frac{\langle\Psi_G|\mathcal{H}|\Psi_G\rangle}{\langle\Psi_G|\Psi_G\rangle} = - \sum_{ij,\sigma} t_{ij}g_{ij}^{(t)} \langle c_{i\sigma}^\dagger c_{j\sigma} \rangle_0  +\text{c.c.}
\nonumber \\
&+&  \sum_{ij} J_{ij}\left( \tfrac{1}{2} g_{ij}^{(s,xy)} \langle S_i^+ S_j^- + S_i^- S_j^+ \rangle_0 + g_{ij}^{(s,z)} \langle S_i^z S_j^z \rangle_0 \right).
\nonumber 
\end{eqnarray}
Here, $\langle \cdots\rangle_0$ denotes expectation values evaluated in the uncorrelated state $\ket{\Psi_0}$. The Gutzwiller factors $g_{ij}$, which renormalize the hopping and spin-exchange processes, encode the effects of strong correlations. 
To describe magnetically ordered states, we use the site-dependent, simplified
form \cite{yang_zhang2009, Christensen2011} of the extended Gutzwiller factors of Ogata and Himeda \cite{ogata_himeda2003}, who generalized the Gutzwiller approximation
to include antiferromagnetic order and nearest-neighbor correlations.

In this scheme, the factors depend on the mean-field parameters, namely the symmetric ($s$) and antisymmetric ($a$) under site exchange, $i\leftrightarrow j$, pairing amplitudes $\Delta_{ij}^{(s)}=\langle c_{i\uparrow}^\dagger c_{j\downarrow}^\dagger -c_{i\downarrow}^\dagger c_{j\uparrow}^\dagger\rangle_0/2$, $\Delta_{ij}^{(a)}=\langle  c_{i\uparrow}^\dagger c_{j\downarrow}^\dagger + c_{i\downarrow}^\dagger c_{j\uparrow}^\dagger \rangle_0/2$, the bond fields $\chi_{ij,\sigma}=\langle c_{i\sigma}^\dagger c_{j\sigma}\rangle_0$, the spin-resolved densities $n_{i\sigma}=\langle c_{i\sigma}^\dagger c_{i\sigma}\rangle_0$ ($n_{i}=n_{i\uparrow}+n_{i\downarrow}$), and the local magnetic moments $m_i=(n_{i\uparrow}-n_{i\downarrow})/2$. More specifically:
\begin{align}
&g_{ij,\sigma}^{(t)} = g_{i,\sigma}^{(t)}g_{j,\sigma}^{(t)},\nonumber\\
&g_{i,\sigma}^{(t)} = \sqrt{\frac{2\delta_i(1-\delta_i)}{1-\delta_i^2+4m_i^2}\frac{1+\delta_i+2\sigma m_i}{1+\delta_i-2\sigma m_i}},\nonumber\\
&g_{ij}^{(s,xy)}=g_{i}^{(s,xy)}g_{j}^{(s,xy)},\nonumber\\
&g_{i}^{(s,xy)}=\frac{2(1-\delta_i)}{1-\delta_i^2+4m_i^2},\nonumber\\
&g_{ij}^{(s,z)}=g_{ij}^{(s,xy)}\frac{2(|\Delta_{ij}^{(s)}|^2 + |\bar{\chi}_{ij}|^2) - 4m_im_jX_{ij}^2}{2(|\Delta_{ij}^{(s)}|^2+|\bar{\chi}_{ij}|^2)-4m_im_j},\nonumber\\
&X_{ij}=1 + \frac{12(1-\delta_i)(1-\delta_j)(|\Delta_{ij}^{(s)}|^2+|\bar{\chi}_{ij}^2|)}{\sqrt{(1-\delta_i^2+4m_i^2)(1-\delta_j^2+4m_j^2)}},
\label{eq_SM:renorm}
\end{align}
where $\sigma=+1(-1)$ for $\uparrow (\downarrow)$, $\delta_i=1-n_i$ denotes the local hole density at site $i$, and $\bar{\chi}_{ij}=\sum_\sigma \chi_{ij,\sigma}/2$ is the spin-averaged bond field. Here $X_{ij}\geq 1$ describes the enhancement of the longitudinal exchange by
antiferromagnetic correlations on the bonds surrounding $\langle i,j\rangle$,
which is associated to the origin of the anisotropy
$g_{ij}^{(s,z)}\neq g_{ij}^{(s,xy)}$~\cite{ogata_himeda2003}.

Our system can therefore be described in terms of a renormalized $t$-$J$ Hamiltonian, whose ground state is represented by the uncorrelated state $\ket{\Psi_0}$, while the effects of strong correlations enter through the Gutzwiller factors. A Hubbard-Stratonovich decoupling of this Hamiltonian can then be performed without discarding these correlation effects. This yields the following mean-field Hamiltonian:
\begin{align*}
    \mathcal{H}_{eff}=&-\sum_{ij,\sigma} t_{ij}g_{ij,\sigma}^{(t)}c_{i\sigma}^\dagger  c_{j\sigma} + h.c.
    \\&- \sum_{ij,\sigma}J_{ij}\left( g_{ij}^{(s,z)}\frac{\chi_{ij,\sigma}^*}{4}+g_{ij}^{(s,xy)}\frac{\chi_{ij,\bar{\sigma}}^*}{2} \right)c_{i\sigma}^\dagger c_{j\sigma}+\text{h.c.}\\
    & -\sum_{ij}J_{ij}\Delta_{ij}^{(s)*}\left( \frac{g_{ij}^{(s,z)}}{4}+ \frac{g_{ij}^{(s,xy)}}{2}\right)(c_{i\uparrow}^\dagger c_{j\downarrow}^\dagger - c_{i\downarrow}^\dagger c_{j\uparrow}^\dagger)+\text{h.c.}\\
    & -\sum_{ij}J_{ij}\Delta_{ij}^{(a)*}\left( \frac{g_{ij}^{(s,z)}}{4}- \frac{g_{ij}^{(s,xy)}}{2}\right)(c_{i\uparrow}^\dagger c_{j\downarrow}^\dagger + c_{i\downarrow}^\dagger c_{j\uparrow}^\dagger)+\text{h.c.}\\
    & +\frac{1}{2}\sum_{ij}J_{ij}g_{ij}^{(s,z)}m_j(c_{i\uparrow}^\dagger c_{i\uparrow} - c_{i\downarrow}^\dagger c_{i\downarrow})+ \text{const.}
\end{align*}
However, since the Gutzwiller renormalization factors depend explicitly on the local mean-field parameters, a direct diagonalization of the mean-field Hamiltonian does not, in general, minimize the full variational energy. To properly account for this dependence, the following effective Hamiltonian is constructed from the first-order variation of the energy with respect to these parameters:
\begin{align}
    \mathcal{H}_\text{eff} &= \sum_{ij, \sigma}\frac{\partial W}{\partial \chi_{ij, \sigma}}c_{i\sigma}^\dagger c_{j\sigma}
    +\frac{1}{2}\sum_{ij}\frac{\partial W}{\partial \Delta_{ij}^{(s)}}(c_{i\uparrow}^\dagger c_{j\downarrow}^\dagger - c_{i\downarrow}^\dagger c_{j\uparrow}^\dagger)+\text{h.c.}
    \nonumber \\
    &+\frac{1}{2}\sum_{ij}\frac{\partial W}{\partial \Delta_{ij}^{(a)}}(c_{i\uparrow}^\dagger c_{j\downarrow}^\dagger + c_{i\downarrow}^\dagger c_{j\uparrow}^\dagger) + \text{h.c.}+\sum_{i, \sigma} \frac{\partial W}{\partial n_{i\sigma}}c_{i\sigma}^\dagger c_{i\sigma}, 
    \nonumber 
\end{align}
where $W=E_\text{var}+\mu\sum_{i}n_i$,
with $\mu$ being the chemical potential.

Explicitly, the derivatives yield:
\begin{widetext}
\begin{align*}
    &\frac{\partial W}{\partial \chi_{ij,\sigma}}=-t_{ij}g_{ij}^{(t)}-J_{ij}\left(\frac{g_{ij}^{(s,z)}}{4}+\frac{g_{ij}^{(s,xy)} }{2}\frac{\chi_{ij,\bar{\sigma}}^*}{\chi_{ij,\sigma}^*}\right)\chi_{ij,\sigma}^*+\left( \frac{\partial W}{\partial \chi_{ij,\sigma}} \right)_g,\\
    & \frac{\partial W}{\partial \Delta_{ij}^{(s)}}= -J_{ij}\left(\frac{g_{ij}^{(s,z)}}{4}+\frac{g_{ij}^{(s,xy)}}{2} \right)\Delta_{ij}^{(s)*}+\left( \frac{\partial W}{\partial \Delta_{ij}^{(s)}} \right)_g,\\
    & \frac{\partial W}{\partial \Delta_{ij}^{(a)}}= -J_{ij}\left(\frac{g_{ij}^{(s,z)}}{4}-\frac{g_{ij}^{(s,xy)}}{2} \right)\Delta_{ij}^{(a)*}+\left( \frac{\partial W}{\partial \Delta_{ij}^{(a)}} \right)_g,\\
    & \frac{\partial W}{\partial n_{i\sigma}} = \mu + \frac{\sigma}{2}\sum_{\langle i,j\rangle}J_{ij} g_{ij}^{(s,z)}m_j  + \left( \frac{\partial W}{\partial n_{i\sigma}} \right)_g,
\end{align*}
\end{widetext}
where ($\partial W/\partial O)_g$ denotes the contribution to $\partial W/\partial O$ arising solely from the dependence of the Gutzwiller $g$ factors on the mean-field $O$. Since $W$ is real-valued, $\partial W/\partial O^*=(\partial W/\partial O)^*$.

\subsection{Self-consistent equations}

For a given wave vector $\mathbf{k}$, the Nambu-Gor'kov Green function in Matsubara-frequency domain space is given by:
\begin{align}
    G(\mathbf{k}, i\omega_n) &= [i\omega_n\mathbb{1}_{4N} - h(\mathbf{k})]^{-1} 
    \nonumber \\
    &=
    \begin{pmatrix}
G_{ee}(\mathbf{k},i\omega_n) & F(\mathbf{k},i\omega_n) \\
 F^\dagger(\mathbf{k},i\omega_n)& G_{hh}(\mathbf{k},i\omega_n)\label{eq: Green}
\end{pmatrix},
\end{align}
with the electronic, hole and anomalous (pairing) blocks being, respectively:
\begin{align*}
    &\left[G_{ee}\right]_{ij,\sigma}(\mathbf{k},i\omega_n)=-\int^\beta_0d\tau e^{i\omega_n\tau}\langle T_\tau c_{\mathbf{k}i\sigma}(\tau) c_{\mathbf{k}j\sigma}^\dagger(0)\rangle,\\
    &[G_{hh}]_{ij,\sigma}(\mathbf{k},i\omega_n)=-\int^\beta_0 d\tau e^{i\omega_n\tau}\langle T_\tau c_{\mathbf{-k}i\sigma}^\dagger(\tau)c_{-\mathbf{k}j\sigma}(0)\rangle,\\
    &[F]_{ij,\sigma}(\mathbf{k},i\omega_n)=-\int^\beta_0d\tau e^{i\omega_n\tau}\langle T_\tau c_{-\mathbf{k}j\bar{\sigma}}(\tau) c_{\mathbf{k}i\sigma}(0)\rangle,
\end{align*}
where $\tau$ denotes imaginary time, $T_\tau$ is the imaginary-time-ordering operator, $\beta$ is the inverse temperature, and $\omega_n$ are the fermionic Matsubara frequencies. The matrix $h(\mathbf{k})$ in \eqref{eq: Green} denotes the effective Hamiltonian in the extended Nambu-spinor basis:
\begin{align*}
    &\mathcal{H}_\text{eff}(\mathbf{k})=\frac{1}{2}\Psi_{\mathbf{k}}^\dagger h(\mathbf{k})\Psi_\mathbf{k}=\frac{1}{2}\Psi_\mathbf{k}^\dagger
\begin{pmatrix}
\xi(\mathbf{k}) & \Delta(\mathbf{k}) \\
\Delta^\dagger(\mathbf{k}) & -\xi^T(-\mathbf{k})
\end{pmatrix}\Psi_\mathbf{k},\\
\end{align*}
with:
\begin{align*}
        \Psi_\mathbf{k} =
\begin{pmatrix}
c_{\mathbf{k}\uparrow} \\
c_{\mathbf{k}\downarrow} \\
c_{-\mathbf{k}\uparrow}^\dagger\\
c_{-\mathbf{k}\downarrow}^\dagger
\end{pmatrix};\ c_{\pm\mathbf{k}\sigma}=
\begin{pmatrix}
c_{\pm\mathbf{k}1\sigma} \\
\vdots \\
c_{\pm\mathbf{k}N\sigma}
\end{pmatrix},
\end{align*}
and $N$ denoting the number of sites in the unit cell (bands). Explicitly, the blocks of $h(\mathbf{k})$ are given by:
\begin{widetext}
\begin{align}
&\xi_{ij}(\mathbf{k})=
\begin{pmatrix}
\partial_{n_{i\up}}W& e^{-i\mathbf{k}\cdot\mathbf{r}_{ij}}\partial_{\chi_{ij,\up}}W & 0 & 0\\
e^{i\mathbf{k}\cdot\mathbf{r}_{ij}}\partial_{\chi_{ji,\up}}W & \partial_{n_{j\up}}W & 0 & 0\\
0 & 0 & \partial_{n_{i\down}}W & e^{-i\mathbf{k}\cdot\mathbf{r}_{ij}}\partial_{\chi_{ij,\down}}W\\
0 & 0 & e^{i\mathbf{k}\cdot\mathbf{r}_{ij}}\partial_{\chi_{ji,\down}}W & \partial_{n_{j\down}}W
\end{pmatrix},\nonumber\\
&\Delta_{ij}(\mathbf{k})=\begin{pmatrix}
0 & 0& 0 & e^{-i\mathbf{k}\cdot\mathbf{r}_{ij}}\left(\partial_{\Delta_{ij}^{(s)}}+\partial_{\Delta_{ij}^{(a)}}\right)W\\
0 & 0 & e^{i\mathbf{k}\cdot\mathbf{r}_{ij}}\left(\partial_{\Delta_{ji}^{(s)}}+\partial_{\Delta_{ji}^{(a)}}\right)W &\\
0& e^{-i\mathbf{k}\cdot\mathbf{r}_{ij}}\left(-\partial_{\Delta_{ij}^{(s)}}+\partial_{\Delta_{ij}^{(a)}}\right)W & 0& 0\\
e^{i\mathbf{k}\cdot\mathbf{r}_{ij}}\left(-\partial_{\Delta_{ji}^{(s)}}+\partial_{\Delta_{ji}^{(a)}}\right)W & 0 & 0& 0
\end{pmatrix}.\label{eq: matrix h(k)}
\end{align}
\end{widetext}
The enlarged basis is required to retain the particle-hole symmetry of the Bogoliubov-de Gennes (BdG) Hamiltonian in the presence of the antisymmetric pairing component  $\tau_x h^*(\mathbf{k})\tau_x = -h(-\mathbf{k})$. Accordingly, each positive-energy eigenstate of $h(\mathbf{k})$ has a negative energy counterpart in $h(-\mathbf{k})$, related by:
\begin{align*}
    \phi_{\mathbf{k}m}^{-}
    &=
    (\tau_x\otimes \mathbb{1}_{2N})\left(\phi_{-\mathbf{k}m}^{+}\right)^*
    =
    \begin{pmatrix}
        \mathbf{v}_{-\mathbf{k}m}^{*}\\
        \mathbf{u}_{-\mathbf{k}m}^*
    \end{pmatrix},
    \quad
    E_{\mathbf{k}m}^{-}
    =
    -E_{-\mathbf{k}m}^{+}.
\end{align*}
Here, we have defined:
\begin{align*}
\mathbf{u}_{\mathbf{k}m}=
\begin{pmatrix}
        \mathbf{u}_{\mathbf{k}m\uparrow}\\
        \mathbf{u}_{\mathbf{k}m\downarrow}
\end{pmatrix}=
\begin{pmatrix}
        u_{\mathbf{k}m1\uparrow}\\
        \vdots\\
        u_{\mathbf{k}mN\uparrow}\\
        u_{\mathbf{k}m1\downarrow}\\
        \vdots\\
        u_{\mathbf{k}mN\downarrow}
\end{pmatrix}; \ \  \mathbf{v}_{\mathbf{k}m}=
\begin{pmatrix}
        v_{\mathbf{k}m1\uparrow}\\
        \vdots\\
        v_{\mathbf{k}mN\uparrow}\\
        v_{\mathbf{k}m1\downarrow}\\
        \vdots\\
        v_{\mathbf{k}mN\downarrow}
\end{pmatrix},
\end{align*}
where $\mathbf{u}_{\mathbf{k}m}$ and $\mathbf{v}_{\mathbf{k}m}$ denote the particle- and hole-sector components, respectively, of the $m$-th BdG eigenvector, with $u_{\mathbf{k}mi\sigma}$ and $v_{\mathbf{k}mi\sigma}$ giving the corresponding amplitudes at site $i$ and spin $\sigma$.

Since our system also preserves inversion symmetry, $\mathcal{P}$, which maps $\mathbf{k}$ onto $-\mathbf{k}$, the particle-hole related eigenergies at fixed $\mathbf{k}$ can be written as $\pm E_{\mathbf{k}}$.

Using these relations, the eigenvectors of $h(\mathbf{k})$ can be arranged into the unitary matrix:
\begin{align*}
    \mathcal{W}_{\mathbf{k}}
    &=
    \begin{pmatrix}
        \phi_{\mathbf{k}1}^{+} & \cdots &
        \phi_{\mathbf{k}2N}^{+} &
        \phi_{\mathbf{k}1}^{-} & \cdots &
        \phi_{\mathbf{k}2N}^{-}
    \end{pmatrix}
    \\
    &=
    \begin{pmatrix}
        \mathbf{u}_{\mathbf{k}1} & \cdots &
        \mathbf{u}_{\mathbf{k}2N} &
        \mathbf{v}_{-\mathbf{k}1}^{*} & \cdots &
        \mathbf{v}_{-\mathbf{k}2N}^{*}\\
        \mathbf{v}_{\mathbf{k}1} & \cdots &
        \mathbf{v}_{\mathbf{k}2N} &
        \mathbf{u}_{-\mathbf{k}1}^{*} & \cdots &
        \mathbf{u}_{-\mathbf{k}2N}^{*}   
    \end{pmatrix}.
\end{align*}
Using this matrix, $G(\mathbf{k}, i\omega_n)$ can be rewritten as:
\begin{align*}
    G(\mathbf{k}, i\omega_n)=\mathcal{W}_\mathbf{k}\begin{pmatrix}
i\omega_n\mathbb{1}_{2N}-\mathbb{E}_{\mathbf{k},2N} & 0 \\
 0 & i\omega_n \mathbb{1}_{2N}+\mathbb{E}_{\mathbf{k},2N}
\end{pmatrix} ^{-1}\mathcal{W}_\mathbf{k}^\dagger,
\end{align*}
where.
$\mathbb{E}_{\mathbf{k},2N}
=\operatorname{diag}(E_{\mathbf{k}1},\ldots,E_{\mathbf{k}2N})$
is the diagonal matrix containing the positive eigenvalues of
$h(\mathbf{k})$. 

By carrying out the block multiplication, one obtains:
\begin{align*}
    &G_{ee}(\mathbf{k},i\omega_n)=\sum_{m>0}\left( \frac{\mathbf{u}_{\mathbf{k}m}\cdot\mathbf{u}_{\mathbf{k}m}^\dagger}{i\omega_n-E_{\mathbf{k}m}}+ \frac{\mathbf{v}_{-\mathbf{k}m}^*\cdot\mathbf{v}_{-\mathbf{k}m}^T}{i\omega_n+E_{\mathbf{k}m}}\right),\\
    &F(\mathbf{k},i\omega_n)=\sum_{m>0}\left( \frac{\mathbf{u}_{\mathbf{k}m}\cdot\mathbf{v}_{\mathbf{k}m}^\dagger}{i\omega_n-E_{\mathbf{k}m}}+ \frac{\mathbf{v}_{-\mathbf{k}m}^*\cdot\mathbf{u}_{-\mathbf{k}m}^T}{i\omega_n+E_{\mathbf{k}m}}\right).
\end{align*}

Upon expressing these Green functions in the imaginary-time domain using the inverse Matsubara transform, one obtains the following set of self-consistent equations (SCE) for the mean-field parameters of the system:
\begin{align*}
    &\left\langle c_{\mathbf{k}i\sigma}^\dagger(0) c_{\mathbf{k}j\sigma}(0^-) \right\rangle=\left[ G_{ee} \right]_{ji,\sigma}(\mathbf{k},0^-)\\
    &=T\sum_{\omega_n,m>0}\left( \frac{u_{\mathbf{k}mj\sigma}u_{\mathbf{k}mi\sigma}^*}{i\omega_n-E_{\mathbf{k}m}}+\frac{v_{-\mathbf{k}mj\sigma}^*v_{-\mathbf{k}mi\sigma}}{i\omega_n+E_{\mathbf{k}m}} \right)e^{i\omega_n0^+},\\
    &\left\langle c_{-\mathbf{k}j\bar{\sigma}}(0) c_{\mathbf{k}i\sigma}(0^-) \right\rangle=\left[ \hat{F} \right]_{ij,\sigma}(\mathbf{k},0^-)\\
    &=T\sum_{\omega_n,m>0}\left( \frac{u_{\mathbf{k}mi\sigma}v_{\mathbf{k}mj\bar{\sigma}}^*}{i\omega_n-E_{\mathbf{k}m}}+\frac{v_{-\mathbf{k}mi\sigma}^*u_{-\mathbf{k}mj\bar{\sigma}}}{i\omega_n+E_{\mathbf{k}m}} \right)e^{i\omega_n0^+}.
\end{align*}
After performing the Matsubara sums, our set of SCE becomes:
\begin{align*}
    \langle c_{i\sigma}^\dagger c_{j\sigma}\rangle=\sum_{\mathbf{k},m>0}&\left[  e^{-i\mathbf{k}\cdot \mathbf{r}_{ij}} u_{\mathbf{k}mi\sigma}^* u_{\mathbf{k}mj\sigma}f(E_{\mathbf{k}m})\right.\\
    &\left.+e^{i\mathbf{k}\cdot \mathbf{r}_{ij}} v_{-\mathbf{k}mi\sigma}v_{-\mathbf{k}mj\sigma}^*(1-f(E_{\mathbf{km)}}) \right],
\end{align*}
\begin{align*}
    \langle c_{i\sigma}^\dagger c_{j\bar{\sigma}}^\dagger\rangle=\sum_{\mathbf{k},m>0}&\left[  e^{-i\mathbf{k}\cdot \mathbf{r}_{ij}} u_{\mathbf{k}mi\sigma}^* v_{\mathbf{k}mj\bar{\sigma}}f(E_{\mathbf{k}m})\right.\\
    &\left.+e^{i\mathbf{k}\cdot \mathbf{r}_{ij}} v_{-\mathbf{k}mi\sigma}u_{-\mathbf{k}mj\bar{\sigma}}^*(1-f(E_{\mathbf{km)}}) \right],
\end{align*}
where $f(x)$ denotes the Fermi-Dirac function.

Starting from an initial guess for the parameters, the Hamiltonian $h(\mathbf{k})$ is constructed and diagonalized. Its eigenergies and eigenstates are then used to evaluate $G(\mathbf{k},0^-)$. Since the relevant matrix elements of the Green function determine the corresponding observables, an updated set of parameters is obtained. This procedure is iterated until self-consistency is reached. 

However, an additional equation is required to update the chemical potential $\mu$ at each iteration while keeping the hole doping $\delta$ fixed. To derive this equation, we rewrite the effective Hamiltonian in terms of the Bogoulibov quasiparticle operators $\gamma_{\mathbf{k}m\sigma}$ ($\gamma_{\mathbf{k}m\sigma}^\dagger$):
\begin{equation}
    \mathcal{H}_\text{eff}=\sum_{\mathbf{k}, m}E_{\mathbf{k}m}\left( \gamma_{\mathbf{k}m}^\dagger \gamma_{\mathbf{k}m}-\frac{1}{2} \right) + \frac{1}{2}\sum_{ i,\sigma}\frac{\partial W}{\partial n_{i\sigma}}.
\end{equation}
The free energy associated with this Hamiltonian reads:
\begin{equation}
    F = -T\sum_{\mathbf{k}, m}\ln\left[ 2\cosh\left( \frac{E_{\mathbf{k}m}}{2T} \right)  \right]+\frac{1}{2}\sum_{i,\sigma}\frac{\partial W}{\partial n_{i\sigma}},
\end{equation}
which is related to the total number of electrons in the system, $N_e$, through the Maxwell relation $N_e=\partial_\mu F$. Using this relation, one finds:
\begin{equation}
    \delta = \frac{1}{2N_kN}\sum_{\mathbf{k}, m}\text{tanh}\left( \frac{E_{\mathbf{km}}}{2T} \right)\frac{\partial E_{\mathbf{k}m}}{\partial \mu},
\end{equation}
where $N_k$ denotes the number of $\mathbf{k}$-points used to discretize the first Brillouin zone (FBZ). The average hole density in the system is defined as $\delta=\sum_{i=1}^{N_B}\delta_i/N$. Applying the Hellmann-Feynman theorem, the previous expression can be written as:
\begin{equation}
    \delta = \frac{1}{2N_k N}\sum_{\mathbf{k}, m}\text{tanh}\left(  \frac{E_\mathbf{km}}{2T} \right)\phi^{+\dagger}_{\mathbf{k}m}\frac{\partial h( \mathbf{k})}{\partial \mu}\phi^+_{\mathbf{k}m},
\end{equation}
where we recall that $\phi_{\mathbf{k}m}^+$ are the eigenstates associated with the positive eigenenergies $E_{\mathbf{k}m}^+$. From the definition of $h(\mathbf{k})$ (\ref{eq: matrix h(k)}) it follows directly that:
\begin{align*}
    \frac{\partial h(\mathbf{k})}{\partial \mu}=
    \begin{pmatrix}
        \mathbb{1}_{2N} & 0\\
        0 & -\mathbb{1}_{2N}
    \end{pmatrix}=\sigma_z\otimes\mathbb{1}_{2N}.
\end{align*}
Therefore, the equation that completes our self-consistent set reads:
\begin{align*}
    \delta=\frac{1}{2N_kN}\sum_{\mathbf{k},m}\text{tanh}\left( \frac{E_{\mathbf{k}m}}{2T} \right)\phi_{\mathbf{k}m}^{+\dagger}(\sigma_z\otimes \mathbb{1}_{2N})\phi^+_{\mathbf{k}m}.
\end{align*}
\subsection{Renormalization factors}

\begin{figure}[h!]
\centering
\includegraphics[width=8.5cm]{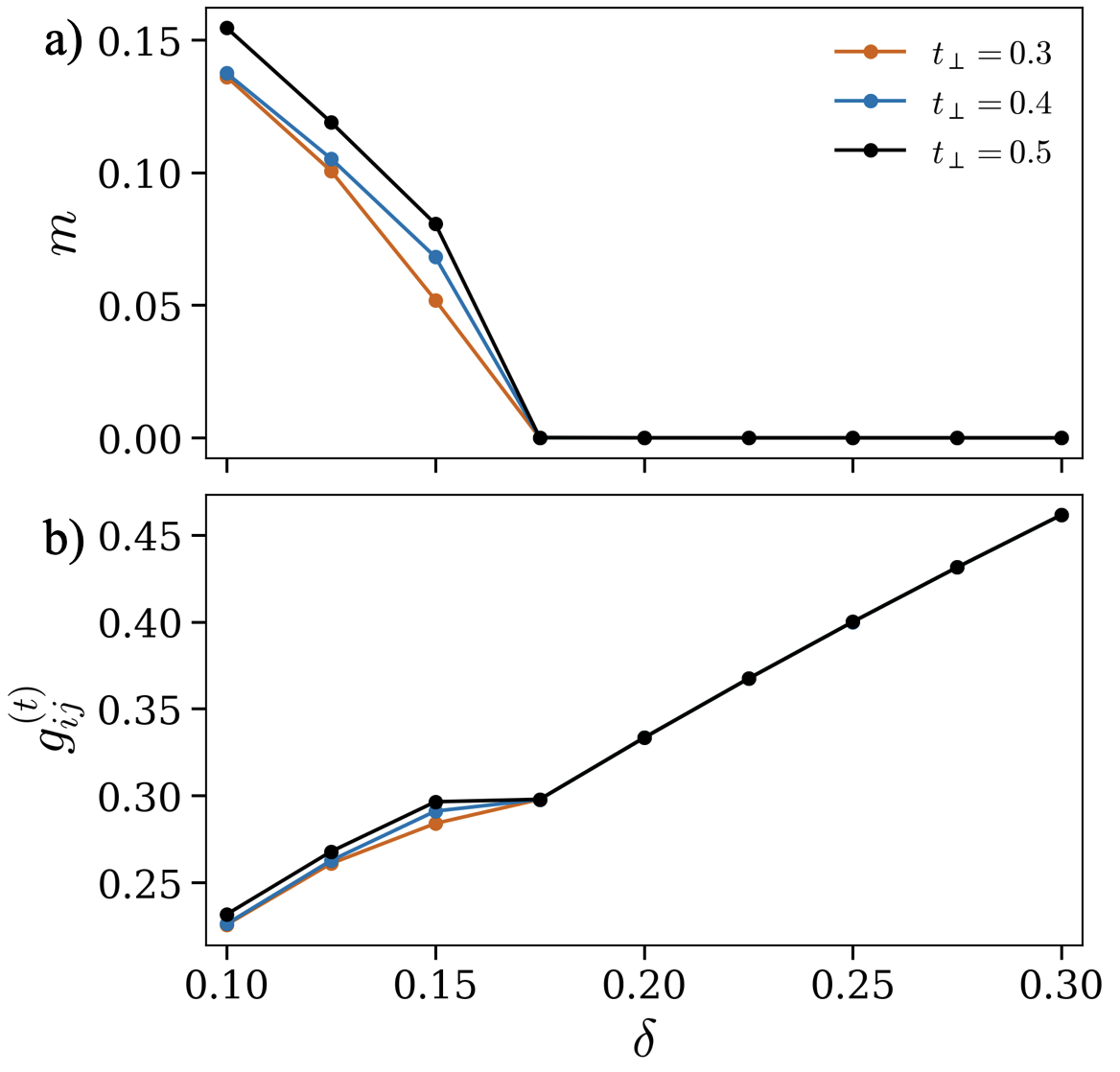}
\caption{Doping dependence of the AFM moment (a) and hopping renormalization factors (b) for a twist angle $\theta =43.6^\circ$, $t'=-0.5t$, and different interlayer couplings $t_\perp$.}
\label{fig:renormalization factors}
\end{figure}

In order to provide further details on the behavior of the renormalization factors in the self-consistent calculations, we focus on the phase diagram of Fig. 2 of the main text, where the AFM moment is gradually suppressed until a transition to the PM state occurs at $\delta_c\sim 0.16-0.17$. In Fig. \ref{fig:renormalization factors}, we show the dependence of the AFM moment $m$ and the hopping renormalization factors $g^{(t)}$ on doping $\delta$ for a twist angle $\theta=43.6^\circ$ and $t'=-0.5t$. Notice that $g_i^{(t)}$ depends only on $m_i$ and $\delta_i$, which exhibit only weak spatial variations arising from differences in the local interlayer coordination of the sites in the twisted geometry. Consequently, all intra- and interlayer hopping renormalization factors are virtually the same. The renormalization factors increase with doping, as expected from the Gutzwiller projection, and display a kink at $\delta_c$ associated with the disappearance of AFM order. For $\delta>\delta_c$, where $m_i=0$, they reduce to the standard Gutzwiller expression within the standard GWA scheme:
\begin{equation}
    g^{(t)}_{ij}=g^{(t)}=\frac{2\delta}{1+\delta}.
\end{equation}

\subsection{s$^{(\pm)}$-wave superconductivity}

When magnetic order is neglected, the introduction of the bonding and antibonding operators:
\begin{align*}
    c_{Bi\sigma}=\frac{c_{1i\sigma}+c_{2i\sigma}}{\sqrt{2}},\ c_{Ai\sigma}=\frac{c_{1i\sigma}-c_{2i\sigma}}{\sqrt{2}},
\end{align*}
allows for the decoupling of the effective Hamiltonian into a sum of a bonding Hamiltonian $H_B$ and an antibonding Hamiltonian $H_A$:
\begin{align*}
H_B &= 
\begin{aligned}[t]
&2\sum_{ij,\sigma}
\frac{\partial W}{\partial \chi_{ij,\sigma}^B}
c^\dagger_{Bi\sigma}c_{Bj\sigma} + 2\sum_{i,\sigma}\frac{\partial W}{\partial n_{i\sigma}^B}c_{Bi\sigma}^\dagger c_{Bi\sigma} \\
&+\sum_{ij}
\frac{\partial W}{\partial \Delta^B_{ij}}
\left(
c^\dagger_{Bi\uparrow}c^\dagger_{Bj\downarrow}
-c^\dagger_{Bi\downarrow}c^\dagger_{Bj\uparrow}
\right)
+\text{h.c.}
\end{aligned}
\\[0.5em]
H_A &= 
\begin{aligned}[t]
&2\sum_{ij,\sigma}
\frac{\partial W}{\partial \chi_{ij,\sigma}^A}
c^\dagger_{Ai\sigma}c_{Aj\sigma} + 2\sum_{i,\sigma}\frac{\partial W}{\partial n_{i\sigma}^A}c_{Ai\sigma}^\dagger c_{Ai\sigma} \\
&+\sum_{ij}
\frac{\partial W}{\partial \Delta^A_{ij}}
\left(
c^\dagger_{Ai\uparrow}c^\dagger_{Aj\downarrow}
-c^\dagger_{Ai\downarrow}c^\dagger_{Aj\uparrow}
\right)
+\text{h.c.}
\end{aligned}
\end{align*}
Here, $n_{i}^{B/A}=(n_{1i}+n_{2i}\pm2\chi_{i}^\perp)/2$, $\chi^{B/A}_{ij,\sigma}=\chi_{ij,\sigma}\pm\chi_{ij,\sigma}^\perp$ and $\Delta^{B/A}_{ij}=\Delta_{ij}\pm\Delta_{ij}^\perp$, with the upper and lower signs corresponding to the bonding and antibonding sectors, respectively.
Since we find $\Delta_{ij}^{(a)}=0$ in all our self-consistent solutions, we denote $\Delta_{ij}^{(s)}=\Delta_{ij}$ to simplify notation.
When $|\Delta_{ij}^\perp|>|\Delta_{ij}|$, which in the aligned bilayer model occurs when the interlayer $s-$wave pairing dominates, both the bonding and antibonding sectors exhibit $s-$wave superconductivity but with opposite signs, $\Delta_{ij}^B\Delta_{ij}^A<0$. Thus, it is this relation between the inter- and intralayer $s-$wave pairing amplitudes that characterizes the $s^\pm-$superconducting state. 
\begin{figure}[t!]
\centering
\includegraphics[width=9cm]{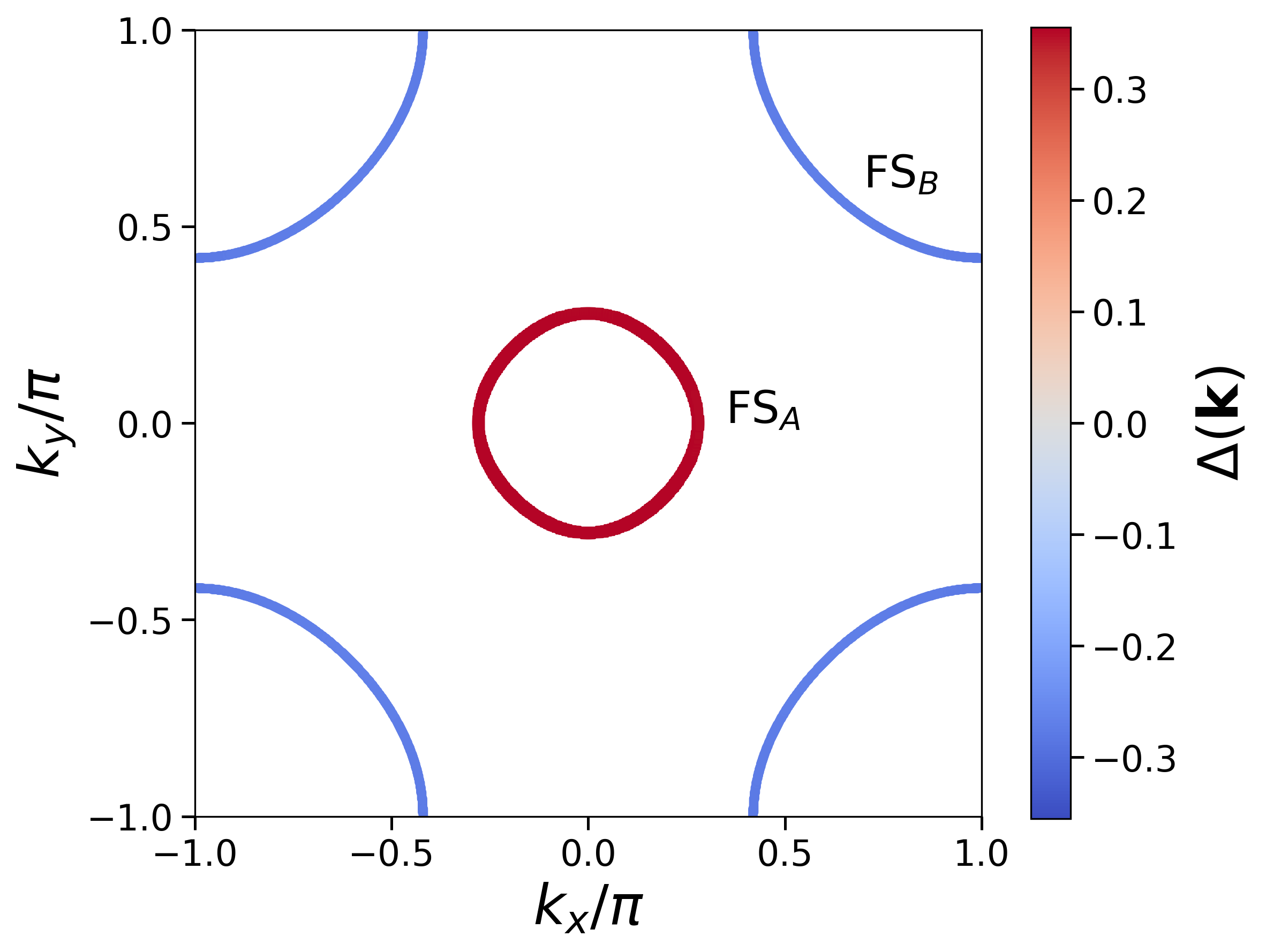}
\caption{Projection of the pairing $\Delta(\mathbf{k})$ onto the bonding ($\text{FS}_B$) and antibonding ($\text{FS}_A$) Fermi surfaces of a paramagnetic homogeneous version of our bilayer system. The opposite signs of the pairing on the two Fermi surfaces provides a clear signature of $s^{(\pm)}-$superconductivity. We take $J=0.3t$, $t'=-0.2t$, and $t_\perp=1.2t$.}
\label{fig: s_pm delta_k}
\end{figure}
To provide further insight into this phase, Fig. \ref{fig: s_pm delta_k} shows the superconducting pairing projected onto the bonding and antibonding Fermi surfaces (FSs) of the homogeneous paramagnetic counterpart of our aligned system. The pairing remains nearly constant along each Fermi surface and has opposite signs in the bonding and antibonding sectors, as expected for an $s^{(\pm)}$-wave superconducting state.

\begin{figure*}[t]
\centering
\includegraphics[width=\textwidth]{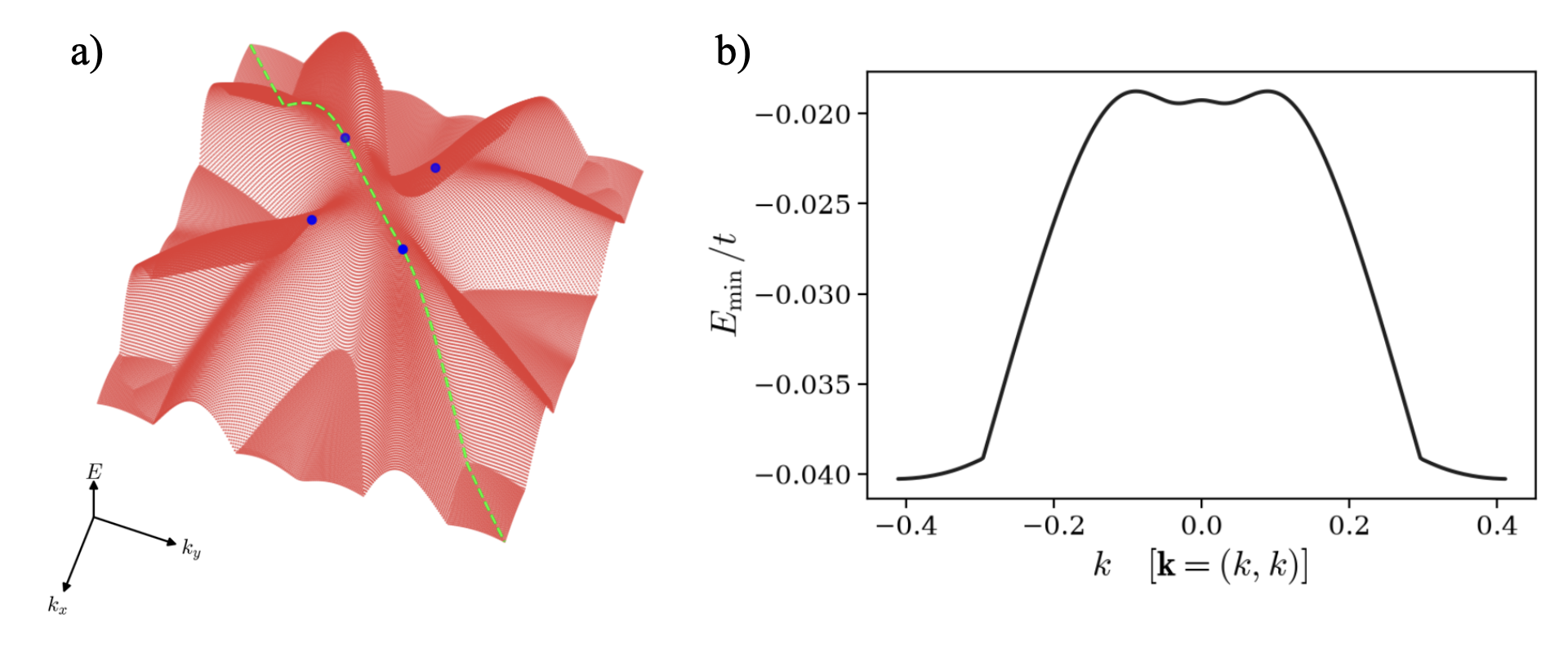}
\caption{Bogoulibov quasiparticle spectrum of the $d_1+id_2$ superconducting phase with $C=2$. (a) Highest negative-energy BdG band across the moiré Brillouin zone for $J=0.3t$, $\delta=0.2$, $t'=-0.5t$ and $t_\perp=0.45t$. The blue markers indicate the four $C_4$-related band maxima, while the green line marks the main diagonal ($k_x=k_y$). (b) Corresponding energy dispersion along this diagonal, showing two maxima associated with the gapped Dirac-like structures }
\label{fig: d_1+id_2 (C=2)}
\end{figure*}
\subsection{$d_1+id_2$ ($C=2$) superconducting phase}

At sufficiently high doping, increasing the interlayer hopping drives the system from the $s+i(d_1+d_2)$ phase into a reentrant $d_1+id_2$ superconducting state but with Chern number $C=2$. This topology can be traced to the nodal quasiparticles of $d-$wave superconductivity, which are gapped by a second $d-$wave component with a relative phase of $\pi/2$. To illustrate this, Fig. \ref{fig: d_1+id_2 (C=2)}(a) shows the highest negative-energy BdG band for $\delta=0.2$, $t'=-0.5t$ and $t_\perp=0.45t$, while Fig. \ref{fig: d_1+id_2 (C=2)}(b) displays its dispersion along the indicated diagonal of the moiré Brillouin zone. The spectrum reveals two prominent gapped Dirac-like structures which, together with their $C_4-$related counterparts highlighted in blue in Fig. \ref{fig: d_1+id_2 (C=2)}(a), suggest the presence of four massive Dirac cones. Within a low-energy Dirac description, each cone can contribute $\pm1/2$ to the Chern number, with the sign determined by its mass and chirality. Four contributions of $+1/2$ would therefore account for the numerically obtained $C=2$.

\subsection{Striped solution}

At a hole doping $\delta=1/8$ we find that a sufficiently negative next-nearest-neighbor hopping $t'$ stabilizes a stripe-ordered state when the two layers are weakly coupled or fully decoupled (see Fig. 1 of the main text). This state exhibits intertwined spin-density-wave (SDW), pair-density-wave (PDW), and charge-density-wave (CDW) orders, coexisting with a uniform $d-$wave superconducting component, consistent with previous studies of $t$-$J$-type models\cite{himeda_ogata2002,yang_zhang2009} on a square lattice.
\begin{figure}[h]
\centering
\includegraphics[width=9cm]{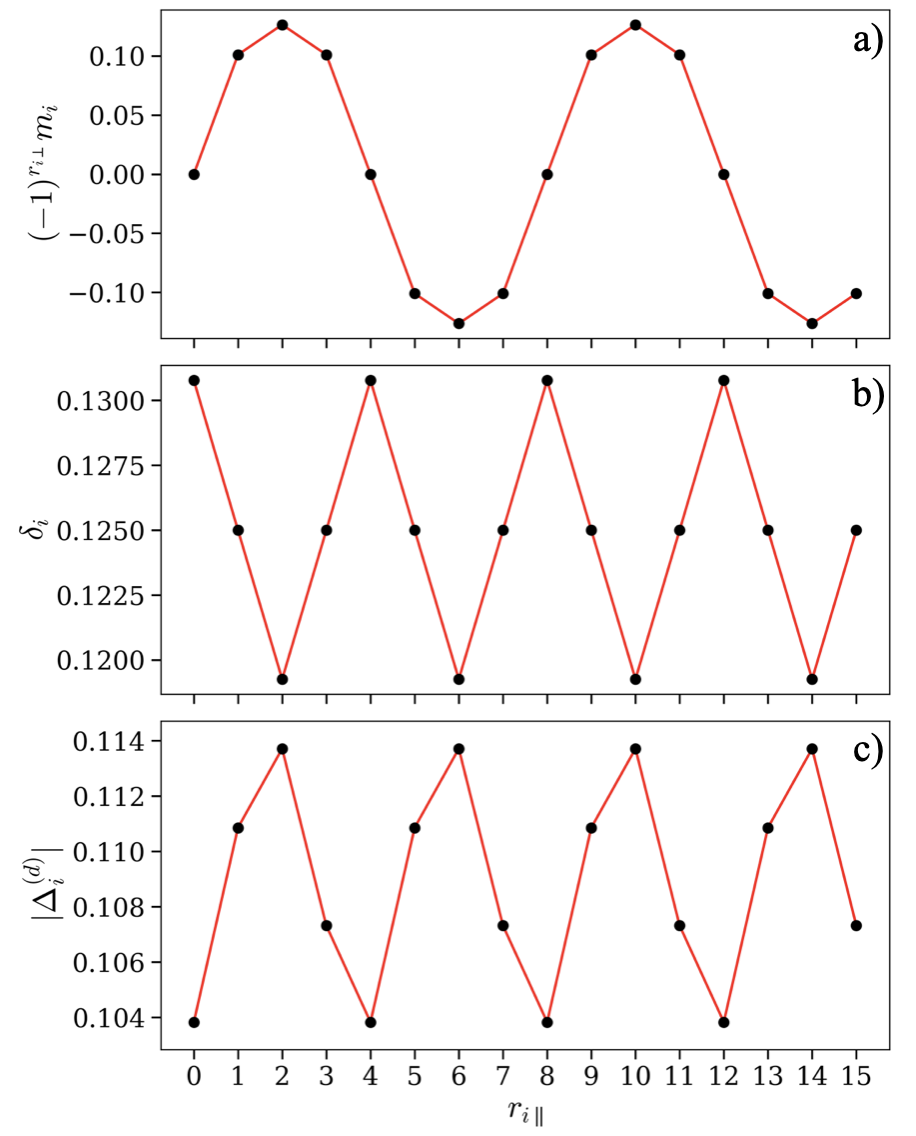}
\caption{Spatial profiles of the staggered magnetization (a), hole density (b), and magnitude of the $d$-wave superconducting pairing order parameter (c) along the direction perpendicular to the stripes, in the fully decoupled limit, $t_\perp=0$ of the $(n,m)=(8, 8)$ bilayer system. The other model parameters are $t'=-0.5t$, $J=0.3t$ and $\delta=0.125$. The staggered magnetization exhibits a periodicity of $\lambda_{m}=1/\delta=8$ lattice sites, whereas the hole density and pairing amplitude display half this period ($\lambda_{\delta,\Delta^{(d)}}=(2\delta)^{-1}=4$). }
\label{fig: stripes}
\end{figure}
The spatial profiles shown in Fig.\ref{fig: stripes} reveal approximately sinusoidal modulations of the self-consistent order parameters along the $r_\perp$ direction, perpendicular to the stripes, of the form:
\begin{align}
m_i&\simeq(-1)^{r_{i\perp}+r_{i\parallel}}m_0\sin (2\pi\delta r_{i\perp}),
\nonumber \\
\Delta_i^{(d)}&\simeq\Delta_0^{(d)}+\Delta_1^{(d)}\cos (4\pi\delta r_{i\perp})
\nonumber \\
\delta_i&\simeq\delta_0+\delta_1\cos(4\pi\delta r_{i\perp}),
\nonumber \\
\end{align}
where $m_i$, $\Delta_i^{(d)}$, and $\delta_i$ denote the local magnetization, $d-$wave pairing amplitude, and hole density, respectively, while $r_{i\parallel}$ ($\r_{i\perp}$) denotes the coordinate of site $i$ along the direction parallel (perpendicular) to the stripes. Note that $\delta$ is the hole doping. The SDW has a modulation period of $\lambda_m=8$ lattice spacings, while the pairing and charge modulations share a period of $\lambda_{\rho,\Delta}=4$. The superconducting order thus combines a uniform $d-$wave component with a finite-momentum PDW modulation.

The hole-rich regions coincide with the suppression of magnetization at the antiphase domain walls separating antiferromagnetically ordered regions. This spatial organization reflects the competition between carrier mobility and magnetic exchange energy, as the accumulation of holes near domain walls facilitates their motion while preserving antiferromagnetic correlations in the surrounding regions \cite{Huang2018}. As the interlayer hopping increases, the uniform antiferromagnetic state becomes energetically favorable over the stripe-ordered solution.

\begin{figure}[htbp]{
           \includegraphics[width=0.5\textwidth]{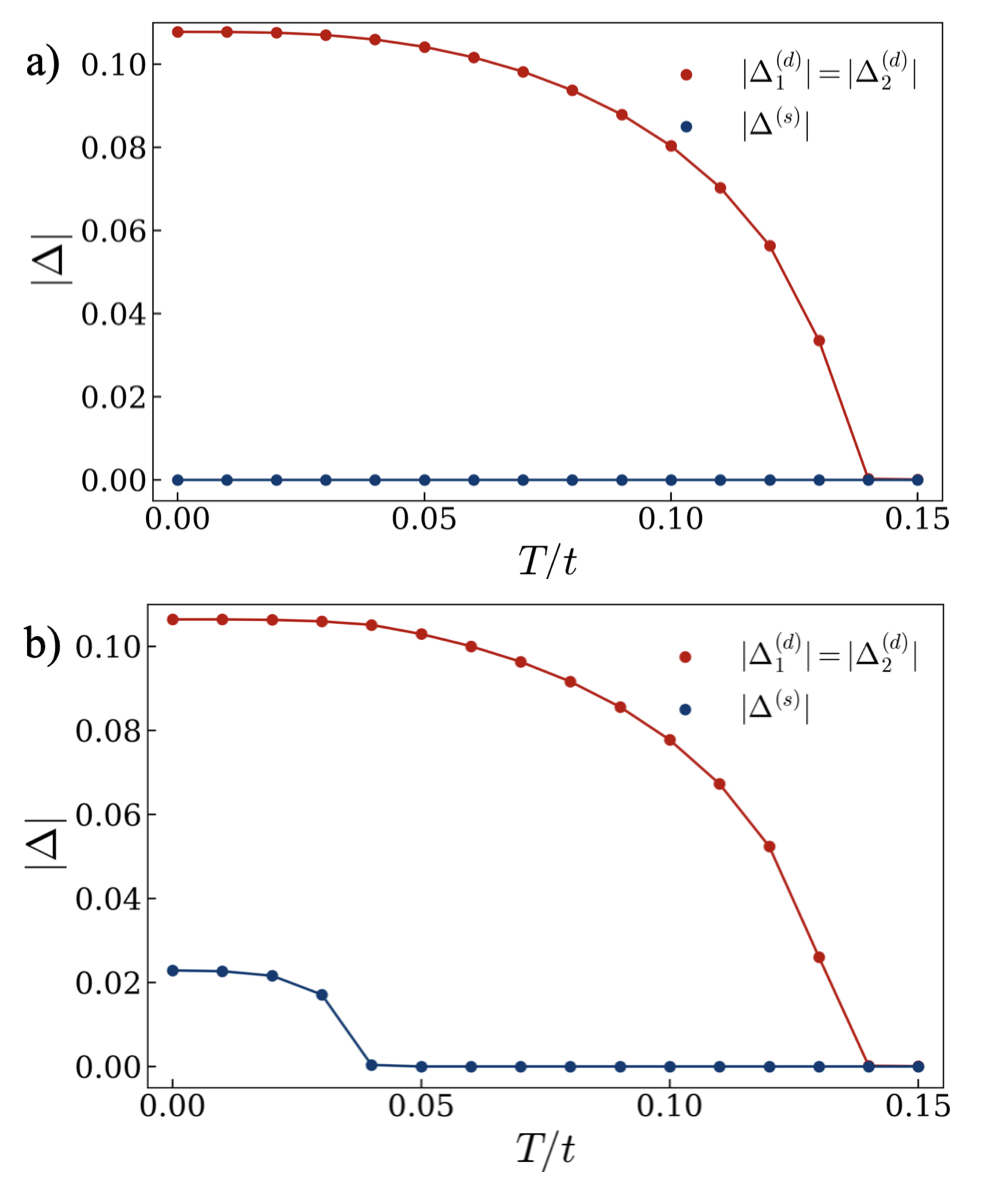}%
    }
    \caption{Temperature dependence 
    of the $d$-wave ($|\Delta^{(d)}_1|=|\Delta^{(d)}_2|$, red) and $s$-wave ($|\Delta^{(s)}|$, blue) gap amplitudes  )(in units of $t$), for a) $d_1+id_2$ pairing at $t_\perp=0.3t$ and b) $s+d_1+id_2$ pairing at  $t_\perp=0.4t$. We have fixed: $\delta=1/8$, $J=0.3t$, and $t'=-0.5t$.
    }
    \label{fig:fig3}
\end{figure}

\subsection{$T$-dependence of superconducting gap}
The $T$-dependence of the gaps, Fig.~\ref{fig:fig3}, shows that the $s$-wave component of the $s+d_1+id_2$ state disappears at $T_s\approx0.04t$, well below the critical temperature $T_c\approx0.14t$ at which the $d$-wave components vanish. This implies two successive transitions on cooling: at $T_c$ the system enters a $C_4$-symmetric (presumably $d_1+id_2$) state, and only below $T_s$ does the $s$-wave admixture set in, lowering the symmetry to $C_2$ and, provided the relative phase of the $d$-wave components remains $\pm\pi/2$, changing the Chern number from $C=4$ to $C=2$. The nematic ARPES pattern and the halving of the number of edge modes discussed below should therefore switch on only below $T_s\ll T_c$, providing an additional, temperature-controlled fingerprint of the $s+d_1+id_2$ state.

\section{Electronic spectral function}
\label{sec:supspectral}

Here we give details on the evaluation of the spectral function relevant 
to ARPES experiments. In Fig. 3 of the main text we show the electronic spectral function $A(\mathbf{k}, \omega)$ for the different time-reversal-breaking superconducting states found.  At zero temperature, the electronic spectral function can be defined through the Lehmann representation as:
{\small
\begin{align}
    A(\mathbf{k},\omega)
    ={}&
    \sum_{i,m,\sigma}
    \left|
    \bra{m,N+1}
    c_{\mathbf{k}i\sigma}^{\dagger}
    \ket{0,N}
    \right|^2
    \delta\left[
        \omega-\left(E_m^{N+1}-E_0^N\right)
    \right]\nonumber
    \\
    &+
    \sum_{i,l,\sigma}
    \left|
    \bra{i,l,N-1}
    c_{\mathbf{k}i\sigma}
    \ket{0,N}
    \right|^2
    \delta\left[
        \omega+\left(E_l^{N-1}-E_0^N\right)
    \right],\nonumber
    \\
    \label{eq: Lehmann}
\end{align}} where $\ket{0,N}$ is the N-particle ground state with energy $E_0^N$, while $\ket{m,N+1}$ and $\ket{l,N-1}$ denote all accessible final states obtained by adding or removing an electron, respectively. Expressing $c_{\mathbf{k}\sigma}$ in terms of the Bogoulibov quasiparticle operators,
{
\begin{align*}
    c_{\mathbf{k}i\sigma}
    =
    \sum_n
    \left[
        u_{i\sigma n}(\mathbf{k})\gamma_{\mathbf{k}n}
        +
        v_{i\sigma n}^{*}(\mathbf{k})
        \gamma_{-\mathbf{k}n}^{\dagger}
    \right],
\end{align*}}
the spectral function can be re-expressed as:
{\small
\begin{align*}
    A(\mathbf{k},\omega)
    =
    -\frac{1}{\pi}
    \lim_{\eta\rightarrow 0^+}
    \operatorname{Im}
    \sum_{i,n,\sigma}
    \left[
        \frac{
            \left|u_{i\sigma n}(\mathbf{k})\right|^2
        }{
            \omega-E_n(\mathbf{k})+i\eta
        }
        +
        \frac{
            \left|v_{i\sigma n}(\mathbf{k})\right|^2
        }{
            \omega+E_n(\mathbf{k})+i\eta
        }
    \right],
\end{align*}}
where $E_n(\mathbf{k})>0$ are the positive BdG eigenergies. In the last expression, the Dirac delta functions appearing in Eq. \eqref{eq: Lehmann}  have been represented as Lorentzians with $\eta \rightarrow 0^+$.

The $A({\bf k},0)$ of the four time-reversal symmetry breaking (TRSB) superconducting states in the phase diagram {\it i.e.} the two $d+id$, the $s+d+id$ and the $s+id$ pairing states are shown in Fig. 3 of the main text. In order to understand the spatial distribution of the spectra, we first discuss the symmetry properties of superconducting gaps under $C_4$ rotations and time reversal, $\mathcal{T}$.  
 In general,  $s + d_1 e^{i \phi_1} + d_2e^{i \phi_2}$ TRSB states ($\phi_1,\phi_2 \neq 0, \pi$) also break the in-plane $C_4$ rotational symmetry
 present in the pure-$d$, $d_1 + id_2$ state~\cite{franz_natphys_2021} remaining $C_2$ symmetric only. While an extra $C_4\mathcal{T}$ symmetry is present in the special $\phi_1, \phi_2= \pm \pi/2$ case, 
it is absent otherwise leading to a $C_2$ nematic superconductor~\cite{Fradkin2010,Fernandes2019,FernandesMillis2013}. 

Thus, while in principle both TRSB $s+i(d_1+d_2)$ and $s+d_1+id_2$ states only preserve $C_2$ symmetry, the breaking of $C_4 \mathcal{T}$ symmetry occurring in the nematic $s+d_1+id_2$ state leads to the fact that it is the only one manifesting $C_2$ symmetry in $A({\bf k},0)$ as shown in Fig. 3 of the main text. This makes ARPES a direct unambiguous probe of this
phase~\cite{Kushnirenko2020}. 

We now explicitly demonstrate on a simple model how the non-nematic $s+i(d_1+d_2)$ state with $C_2$ and $C_4 \mathcal{T}$ symmetries displays $C_4$ symmetry in $A({\bf k},0)$. We consider a single band Bogoliubov-de Gennes (BdG) hamiltonian $h(\mathbf{k})$:
\begin{equation}
h(\mathbf{k}) = \left(
\begin{array}{cccc}
\varepsilon_\uparrow(\mathbf{k}) & 0 & 0 & \Delta(\mathbf{k}) \\
0 & \varepsilon_\downarrow(\mathbf{k}) & -\Delta(\mathbf{k}) & 0 \\
0 & -\Delta^*(\mathbf{k}) & -\varepsilon_\uparrow(-\mathbf{k}) & 0 \\
\Delta^*(\mathbf{k}) & 0 & 0 & -\varepsilon_\downarrow(-\mathbf{k}),
\end{array}
\right),
\\
\label{eq:bdg}
\end{equation}
defined on the four-component extended spinor $\Psi_{\mathbf{k}} = \begin{pmatrix} c_{\mathbf{k}\uparrow} & c_{\mathbf{k}\downarrow} & c^\dagger_{-\mathbf{k}\uparrow} & c^\dagger_{-\mathbf{k}\downarrow} \end{pmatrix}^T$.

We now find the conditions for $C_4 \mathcal{T}$-symmetry with $\mathcal{T} = \mathcal{U}_T \mathcal{K}$, where $\mathcal{K}$ represents the complex conjugation operator and $\mathcal{U}_T
= \mathbb{1}_{2} \otimes i\sigma_y$. $C_4$
transforms, $R_{C_4}:(k_x,k_y) \rightarrow (k_y, -k_x)$.
Applying $C_4 \mathcal{T}$ to the hamiltonian:

\begin{widetext}
\begin{align}
&R_{C_4}\mathcal{T} \left( G^{-1}(\mathbf{k}, i\omega_n) \right) \mathcal{T}^{-1} R_{C_4}^{-1} 
= \mathcal{U}_T \left( -i\omega_n \mathbb{1}_{4\times4} - h^*(-\mathbf{k}) \right)\mathcal{U}_T^{-1} \nonumber \\
&= -i\omega_n \mathbb{1}_{4\times4} -\begin{pmatrix}
\varepsilon_\downarrow(-R_{C_4}\mathbf{k}) & 0 & 0 & \Delta^*(-R_{C_4}\mathbf{k}) \\
0 & \varepsilon_\uparrow(-R_{C_4}\mathbf{k}) & -\Delta^*(-R_{C_4}\mathbf{k}) & 0 \\
0 & -\Delta(-R_{C_4}\mathbf{k}) & -\varepsilon_\downarrow(R_{C_4}\mathbf{k}) & 0 \\
\Delta(-R_{C_4}\mathbf{k}) & 0 & 0 & -\varepsilon_\uparrow(R_{C_4}\mathbf{k})
\end{pmatrix} 
\end{align}
\end{widetext}
Thus, the $C_4 \mathcal{T}$ symmetry is automatically satisfied when:
\begin{equation}
\label{eq:TRS_conditions}
\begin{aligned}
\varepsilon_\uparrow(-R_{C_4}\mathbf{k}) &= \varepsilon_\downarrow(\mathbf{k}), \\
\varepsilon_\downarrow(-R_{C_4}\mathbf{k}) &= \varepsilon_\uparrow(\mathbf{k}), \\
\Delta^*(-R_{C_4}\mathbf{k}) &= \Delta(\mathbf{k}).
\end{aligned}
\end{equation}
In the present simple model we take the relevant spin degenerate single band: $\epsilon_\sigma(\mathbf{k})=\epsilon({\bf k})=-2 t (\cos(k_x)+\cos(k_y))$ with the $s+d_1+i d_2$ gap:
\begin{eqnarray}
\Delta(\mathbf{k}) &=& \Delta_s + i \Delta_{d_1} ( \cos k_x - \cos k_y) 
\nonumber \\
&+& i \Delta_{d_2}\sin k_x \sin k_y,
\nonumber \\
\end{eqnarray}
automatically satisfying the above $C_4 \mathcal{T}$-symmetry conditions  \eqref{eq:TRS_conditions}. Note that the gap is not invariant under ${C_4}$ nor complex conjugation acting separately.
So the $s + i(d_1 + d_2)$ superconductor breaks $\mathcal{T}$ and $C_4$ symmetries but preserves $\mathcal{T} C_4$ symmetry.
This is because the $C_4$ transformation of the gap is equivalent to the complex conjugation in \eqref{eq:TRS_conditions}.

The spectral density associated with $h({\bf k})$ above can be obtained from the electron diagonal elements of the electron part, $G^e({\bf k}, \omega+i \eta)$, as:
\begin{eqnarray}
A_\sigma({\bf k}, \omega) &=& -{1 \over \pi} \text{Im}[G_{ee}]_{ii}({\bf k},\omega+i \eta) 
\nonumber \\
&=& {1 \over \pi} \text{Im} \left\{ { \epsilon({\bf k} )+\omega +i \eta \over \epsilon({\bf k})^2+| \Delta({\bf k})|^2 + \eta^2 - \omega^2 - i 2 \eta \omega} \right\}
\nonumber \\
&=& {1 \over \pi} { \eta ( \epsilon({\bf k} )^2+\omega^2 + |\Delta({\bf k})|^2+\eta^2 ) +2 \eta \omega \epsilon \over (\omega^2 -\eta^2-\epsilon({\bf k})^2 -|\Delta({\bf k})|^2)^2 + 4 \eta^2 \omega^2}.
\label{eq:akw}
\end{eqnarray}
Crucially, since the diagonal elements of the Nambu Greens function, $G^e_{\sigma\sigma}({\bf k, \omega})$ and thus, $A_\sigma({\bf k}, \omega)$ depend on the gap modulus, $|\Delta({\bf k})|^2=|\Delta_s|^2 + |\Delta_{d_1}+\Delta_{d_2}|^2$, 
$A_\sigma({\bf k}, \omega)$  is invariant under $C_4$ even though the full $G({\bf k}, \omega)$ is not. This is again due to the  $C_4 \mathcal{T}$ symmetry satisfied by the $s+i(d_1+d_2)$ superconductor even though it breaks $C_4$ but just as a complex conjugation transformation.  An example of the $\omega=0$ spectra on a square 
lattice is given in Fig. \ref{fig:SM_berry} (a).

Consequently, the $A_\sigma(\mathbf{k},\omega)$ of the $s+i(d_1 + d_2)$ state is $C_4$ invariant despite the lower $C_2$ symmetry of the gap. This is consistent with the vanishing off-diagonal optical conductivities, $\sigma_{xy}=\sigma_{yx}=0$, reported for these states ~\cite{li_arxiv_2026}, whereas the $C_2$ pattern of the $s+d_1+id_2$ phase ( $\sigma_{xy}+\sigma_{yx}\neq0$) directly reflects its nematic order. In Sec. \ref{sec:SM_chern} we demonstrate how the $C_4\mathcal{T}$ symmetry of the $s+i(d_1 + d_2)$ state implies a zero Chern number, $C=0$. The absence of such symmetry in $d_1+id_2$ and nematic $s+d_1e^{i\phi_1}+d_2e^{i\phi_2}$ phases leads to a finite Chern number which can be calculated explicitly.
 
We finally introduce an extension of the previous model to analyze $A_\sigma({\bf k{, \omega}})$ when $s+i(d_1+d_2)$ superconductivity coexists with AFM order relevant to the low doping regime 
of our phase diagrams (see Fig. 1 and Fig. 2 in the 
main text). We introduce an extended 8-component spinor to include the two-sublattice AFM order as well as the superconducting phases: 
\begin{equation}
\Psi_{\mathbf{k}} = \begin{pmatrix} c_{0,\mathbf{k}\uparrow} & c_{1,\mathbf{k}\uparrow} &c_{0,\mathbf{k}\downarrow} & c_{1,\mathbf{k}\downarrow} & c^\dagger_{0,-\mathbf{k}\uparrow} & c^\dagger_{1,-\mathbf{k}\uparrow} & c^\dagger_{0,-\mathbf{k}\downarrow} & c^\dagger_{1,-\mathbf{k}\downarrow} \end{pmatrix}^T.
\end{equation}
with the corresponding BdG hamiltonian:
\begin{widetext}
\begin{equation}
h_\text{AFM}({\bf k}) = 
\begin{pmatrix}
m + \mu & t_{AB}({\bf k}) & 0 & 0 & 0 & 0 & 0 & \frac{1}{2}\Delta({\bf k}) \\
t_{AB}({\bf k}) & -m + \mu & 0 & 0 & 0 & 0 & \frac{1}{2}\Delta({\bf k}) & 0 \\
0 & 0 & -m + \mu & t_{AB}({\bf k}) & 0 & -\frac{1}{2}\Delta({\bf k}) & 0 & 0 \\
0 & 0 & t_{AB}({\bf k}) & m + \mu & -\frac{1}{2}\Delta({\bf k}) & 0 & 0 & 0 \\
0 & 0 & 0 & -\frac{1}{2}\Delta^*({\bf k}) & -m - \mu & -t_{AB}({\bf k}) & 0 & 0 \\
0 & 0 & -\frac{1}{2}\Delta^*({\bf k}) & 0 & -t_{AB}({\bf k}) & m - \mu & 0 & 0 \\
0 & \frac{1}{2}\Delta^*({\bf k}) & 0 & 0 & 0 & 0 & m - \mu & -t_{AB}({\bf k}) \\
\frac{1}{2}\Delta^*({\bf k}) & 0 & 0 & 0 & 0 & 0 & -t_{AB}({\bf k}) & -m - \mu
\end{pmatrix}.
\end{equation}
\end{widetext}
where $m$ is the local moment on each site. Due to the AFM order the $C_4\mathcal{T}$ symmetry is now broken (apart from breaking $C_4$ and $\mathcal{T}$ separately) but $C_4U_y(-\pi)\mathcal{T}$ is satisfied. This is equivalent to a $C_4\mathcal{K}$ transformation\cite{Chen2025}: $R_{C_4}\mathcal{K}h_\text{AFM}(k)\mathcal{K}^{-1}R_{C_4}^{-1} \equiv h_\text{AFM}^*(-R_{C_4}{\bf k})$, which does not include a spin inversion. One finds that $h_{AFM}({\bf k})$ is invariant under the combined $C_4\mathcal{K}$ transformation on the matrix elements: 
$h_\text{AFM}^*(-R_{C_4}{\bf k})=h_\text{AFM}(k)$ since $h_\text{AFM}({\bf k})$ is hermitian and $\Delta^*(-R_{C_4}\mathbf{k}) = \Delta(\mathbf{k})$ in the $s+i(d_1+d_2)$ state. 

The electronic spectral function obtained from $h_\text{AFM}({\bf k})$ reads:
\begin{equation}
A_{s\sigma}({\bf k}, \omega)=-{1 \over \pi} \text{Im}{ -(\omega+ (-1)^s\sigma m) -i \eta \over -(i \eta +\omega)^2+ m^2+t_{AB}({\bf k})^2 + {|\Delta({\bf k})|^2\over 4 } }
\end{equation}
where $s=0,1$ denotes the sublattice. 
Since $A_{s\sigma}$ depends on the gap only through its modulus $|\Delta({\bf k})|$, it remains invariant under a $C_4$ rotation
even though the $s+i(d_1+d_2)$ gap is only $C_2$ rotationally invariant. 
Note that the off-diagonal anomalous terms of the Nambu-Gor'kov Green function are, in contrast, proportional to $\Delta(\mathbf{k})$ and so would break $C_4$ symmetry reflecting the actual $C_2$ symmetry of the pairing state.

\section{Vanishing Chern number of the \texorpdfstring{$s+i(d_1+d_2)$}{s+i(d1+d2)} state}
\label{sec:SM_chern}
 
As mentioned above, among the four TRSB superconducting states encountered, the $s+i(d_1+d_2)$ is the only one with vanishing Chern number. Here we demonstrate that its $C_2$ and $C_4\mathcal{T}$ symmetries naturally enforces $C=0$, as previously discussed in the context of $C_4\mathcal{T}$-symmetric magnetic insulators~\cite{GonzalezHernandez2025}. 
 
To make the derivation as simple as possible we now consider the $2 \times 2$ BdG hamiltonian: 
\begin{equation}
h(\mathbf{k})=\begin{pmatrix}\epsilon({\mathbf{k}}) & \Delta({\mathbf{k}})\\ 
\Delta^*(\mathbf{k}) & -\epsilon(\mathbf{k})
\end{pmatrix}=\mathbf{d}(\mathbf{k})\cdot\boldsymbol{\tau},
\label{eq:SM_HBdG}
\end{equation}
defined on the $\Psi_{\mathbf{k}}=(c_{\mathbf{k}\uparrow},\,c^\dagger_{-\mathbf{k}\downarrow})^T$ spinor. The $\boldsymbol{\tau}$ are Pauli matrices in particle--hole space, $\mathbf{d}=(\Delta^R({\bf k}),\,-\Delta^I({\bf k}),\,\epsilon({\bf k}))$ with Bogoliubov quasiparticle energies $E_\pm=\pm|\mathbf{d}|$. In the $s+i(d_1+d_2)$ state we have: 
\begin{eqnarray}
\Delta^R({\bf k}) &=&\Delta_s
\nonumber \\
\Delta^I({\bf k}) &=& \Delta_{d_1} (\cos(k_x)-\cos(k_y)) +\Delta_{d_2} \sin(k_x) \sin(k_y) 
\nonumber \\
\end{eqnarray}
From the discussion in the previous section this hamiltonian is $C_4 \mathcal{T}$-symmetric since the $s+i(d_1+d_2)$ state  satisfies the conditions \eqref{eq:TRS_conditions}.  

 \begin{figure*}[t]
\centering
\includegraphics[width=\textwidth]{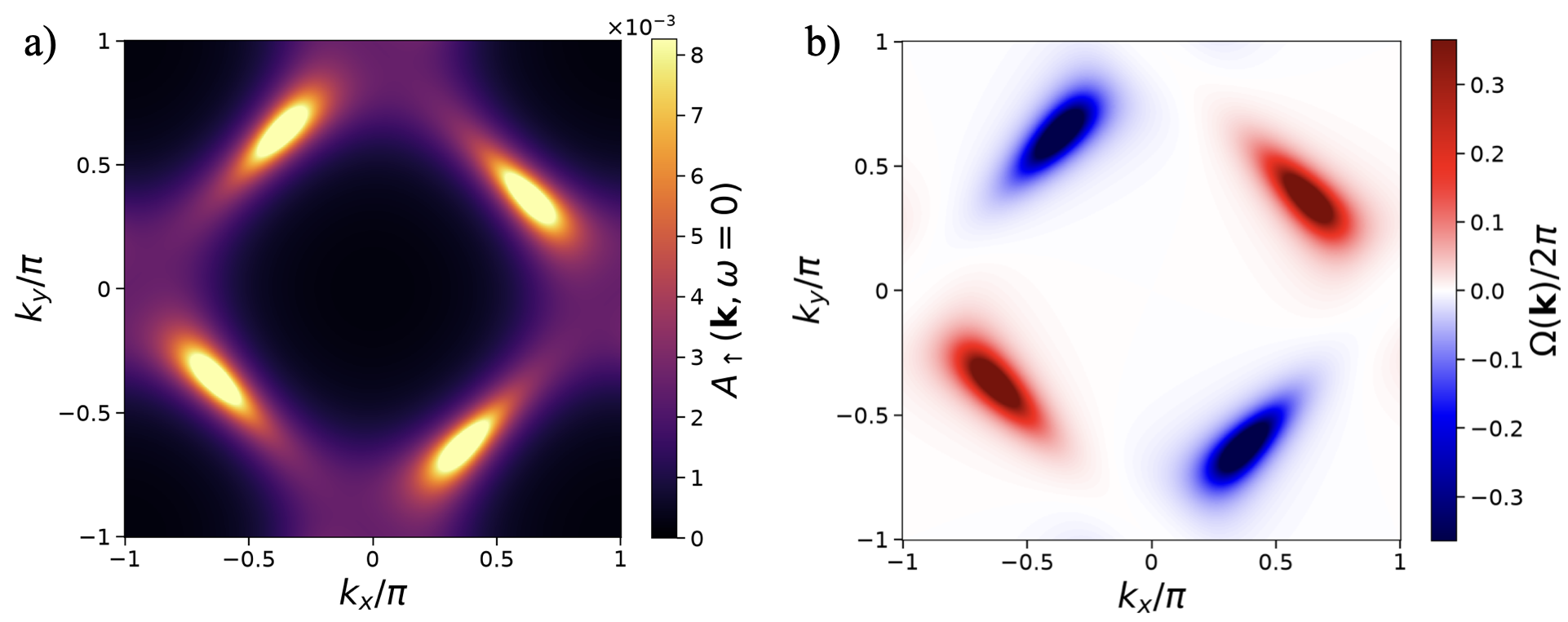}
\caption{(a) Spin resolved spectral function at the Fermi energy, $A_{\sigma}({\bf k,\omega})$, of the $s+i(d_1+d_2)$ state in the first Brillouin
zone of the square lattice. Note how the spectra exhibits $C_4$ rotational invariance although the  $s+i(d_1+d_2)$ state does not.
(b) Corresponding Berry curvature $\Omega(\mathbf{k})$ adding up to a zero Chern number, $C=0$.  
We consider simple model \eqref{eq:bdg}  with $t'=0$, broadening $\eta=0.01$, and $\Delta_s = \Delta_{d_1} =\Delta_{d_2}=0.5$) in these plots.}
\label{fig:SM_berry}
\end{figure*}
We can now obtain the Berry curvature and Chern number from 
the occupied (negative-energy) states $|u(\mathbf{k})\rangle$:
\begin{align}
\Omega(\mathbf{k})&=i\big(\langle\partial_{k_x}u|\partial_{k_y}u\rangle-\langle\partial_{k_y}u|\partial_{k_x}u\rangle\big),
\label{eq:SM_Omega}\\
C&=\frac{1}{2\pi}\int_{\rm BZ}\Omega(\mathbf{k})\,d^2k\in\mathbb{Z}.
\label{eq:SM_C}
\end{align}
This expression is numerically evaluated \cite{fukui2005} in the various TRSB superconducting states found in the paper. 
For the simple Hamiltonian~\eqref{eq:SM_HBdG} discussed here, the Berry curvature takes the form~\cite{QWZ2006}:
\begin{equation}
\Omega(\mathbf{k})=\frac{1}{2|\mathbf{d}|^3}\,\mathbf{d}\cdot\big(\partial_{k_x}\mathbf{d}\times\partial_{k_y}\mathbf{d}\big),
\label{eq:SM_twoband}
\end{equation}
and with $\mathbf{d}=(\Delta^R({\bf k}),-\Delta^I({\bf k}),\epsilon({\bf k}))$ and the Jacobian bracket $\{f,g\}\equiv\partial_{k_x}f\,\partial_{k_y}g-\partial_{k_y}f\,\partial_{k_x}g$, the triple product reads
\begin{eqnarray}
\mathbf{d}\cdot(\partial_{k_x}\mathbf{d}\times\partial_{k_y}\mathbf{d}) = &-&\Delta^R({\bf k})\{\Delta^I({\bf k}),\epsilon({\bf k})\}\nonumber\\ 
&+&\Delta^I({\bf k})\{\Delta^R({\bf k}),\epsilon({\bf k})\}
\nonumber \\
&-&\epsilon({\bf k})\{\Delta^R({\bf k}),\Delta^I({\bf k})\}.
\label{eq:SM_triple}
\end{eqnarray}
Two properties follow. First, the bracket is linear in each argument and therefore transforms like the product $fg$ under $\Delta^I({\bf k})\to -\Delta^I({\bf k})$; since $\Delta^I({\bf k})$ enters each term of Eq.~\eqref{eq:SM_triple} exactly once, either undifferentiated or inside the bracket, the numerator of Eq.~\eqref{eq:SM_twoband} is odd in $\Delta^I$, while $|\mathbf{d}|^3=(\Delta^R({\bf k})^2+\Delta^I({\bf k})^2+\epsilon({\bf k})^2)^{3/2}$ is even. Writing $\Omega[\Delta^R,\Delta^I,\epsilon]$ for the curvature built from these functions:
\begin{equation}
\Omega[\Delta^R({\bf k}),-\Delta^I({\bf k}),\epsilon({\bf k})]=-\,\Omega[\Delta^R({\bf k}),\Delta^I({\bf k}),\epsilon({\bf k})],
\label{eq:SM_oddB}
\end{equation}
{\it i.e.}, the curvature changes sign under complex conjugation of $h({\bf k})$ ($ \Delta^I({\bf k})\rightarrow -\Delta^I({\bf k})$). Second, a $C_4$ rotation leaves the brackets unchanged: for the rotated functions $\tilde f(\mathbf{k})\equiv f(C_4\mathbf{k})$, the chain rule gives $\nabla\tilde f(\mathbf{k})=C_4^T(\nabla f)(C_4\mathbf{k})$, explicitly $(-\partial_{k_y} f,\partial_{k_x} f)|_{C_4\mathbf{k}}$, and the two minus signs cancel in the bracket ($\det C_4=1$), so $\{\tilde f,\tilde g\}(\mathbf{k})=\{f,g\}(C_4\mathbf{k})$ and $\Omega[\Delta^R,\Delta^I,\epsilon](C_4\mathbf{k})=\Omega[\tilde \Delta^R,\tilde \Delta^I,\tilde\epsilon](\mathbf{k})$. Since $h(C_4\mathbf{k})=h(\mathbf{k})^*$, $\tilde \Delta^R({\bf k})=\Delta^R({\bf k})$, $\tilde \Delta^I({\bf k})=-\Delta^I({\bf k})$ and $\tilde\epsilon({\bf k})=\epsilon({\bf k})$, so that, with Eq.~\eqref{eq:SM_oddB},
\begin{equation}
\Omega(C_4\mathbf{k})=\Omega[\Delta^R({\bf k}),-\Delta^I({\bf k}),\epsilon(\mathbf{k})]=-\,\Omega(\mathbf{k}).
\label{eq:SM_Omodd}
\end{equation}
Finally, the substitution $\mathbf{k}'=C_4\mathbf{k}$ maps the square BZ $[-\pi,\pi]^2$ onto itself with unit Jacobian, $d^2k'=d^2k$, and \eqref{eq:SM_C} give
\begin{equation}
\begin{aligned}
2\pi C&=\int_{\rm BZ}\Omega(\mathbf{k}')\,d^2k'=\int_{\rm BZ}\Omega(C_4\mathbf{k})\,d^2k\\
&=-\int_{\rm BZ}\Omega(\mathbf{k})\,d^2k=-2\pi C,
\end{aligned}
\label{eq:SM_Cint}
\end{equation}
hence $C=0$, necessarily. The curvature does not vanish, but its contributions from regions related by $90^\circ$ rotations cancel in pairs, as illustrated in Fig.~\ref{fig:SM_berry}(a) for $\Delta_s=\Delta_0=0.5$. 

Now we can understand how no symmetry constraint of the type discussed is imposed on the $\Omega(\mathbf{k})$ of the other TRSB superconducting states leading to finite $C$. For $d_1+id_2$, $\Delta^R=\Delta_{d_1}(\cos k_x-\cos k_y)$ and  $\Delta^I=\Delta_{d2}\sin k_x \sin k_y$ both components are odd under $C_4$, which therefore maps $\Delta({\bf k})\to-\Delta({\bf k})$ (a global $U(1)$ gauge phase). Each term in Eq.~\eqref{eq:SM_triple} contains a $\Delta^R({\bf k})$ and one $\Delta^I({\bf k})$, so $\Omega(C_4\mathbf{k})=\Omega(\mathbf{k})$ and the symmetry imposes no constraint. For instance, for the $s+d_1+id_2$, $\Delta^R({\bf k})=\Delta_s+\Delta_{d1}(\cos k_x - \cos k_y)$, 
has no definite parity under $C_4$, so no constraint arises as expected.

An alternative route follows from the antiunitary symmetry and the definition~\eqref{eq:SM_Omega} alone. We have seen how invariance under $C_4\mathcal{T}$ in the $s+i(d_1+d_2)$ implies $h(C_4\mathbf{k})=h(\mathbf{k})^*$. If $h(\mathbf{k})|u(\mathbf{k})\rangle=E|u(\mathbf{k})\rangle$, complex conjugation gives $h(C_4\mathbf{k})|u(\mathbf{k})\rangle^*=E\,|u(\mathbf{k})\rangle^*$, so that, up to a phase, the occupied state at $C_4\mathbf{k}$ is the complex conjugate of that at $\mathbf{k}$:
\begin{equation}
|u(C_4\mathbf{k})\rangle=e^{i\chi(\mathbf{k})}\,|u(\mathbf{k})\rangle^* .
\label{eq:SM_map}
\end{equation}
Consider the Berry curvature of $|v(\mathbf{k})\rangle\equiv|u(C_4\mathbf{k})\rangle$, i.e.\ Eq.~\eqref{eq:SM_Omega} with $u\to v$, and evaluate it in two ways. First, by rotation: under $C_4, {\bf k}$ $\partial_{k_x}\rightarrow-\partial_{k_y}$ and $\partial_{k_y}\rightarrow\partial_{k_x}$, and the two minus signs cancel in the antisymmetric combination~\eqref{eq:SM_Omega}, so $\Omega_v(\mathbf{k})=\Omega(C_4\mathbf{k})$; since $d^2q=d^2k$ and the BZ is mapped onto itself, $\int_{\rm BZ}\Omega\,d^2k=2\pi C$. Second, by antiunitarity: according to Eq.~\eqref{eq:SM_map}, $|v\rangle$ is $|u\rangle^*$ up to a phase, which drops out, and complex conjugation exchanges the two terms in Eq.~\eqref{eq:SM_Omega}, since $\langle\partial_{k_x}u^*|\partial_{k_y}u^*\rangle=\langle\partial_{k_x}u|\partial_{k_y}u\rangle^*=\langle\partial_{k_y}u|\partial_{k_x}u\rangle$; hence $\Omega_v(\mathbf{k})=-\Omega(\mathbf{k})$ and $\int_{\rm BZ}\Omega_v\,d^2k=-2\pi C$. Comparing both, $C=-C=0$, and pointwise $\Omega(C_4\mathbf{k})=-\Omega(\mathbf{k})$, Eq.~\eqref{eq:SM_Omodd}. 

 \begin{figure*}[t]
\centering
\includegraphics[width=\textwidth]{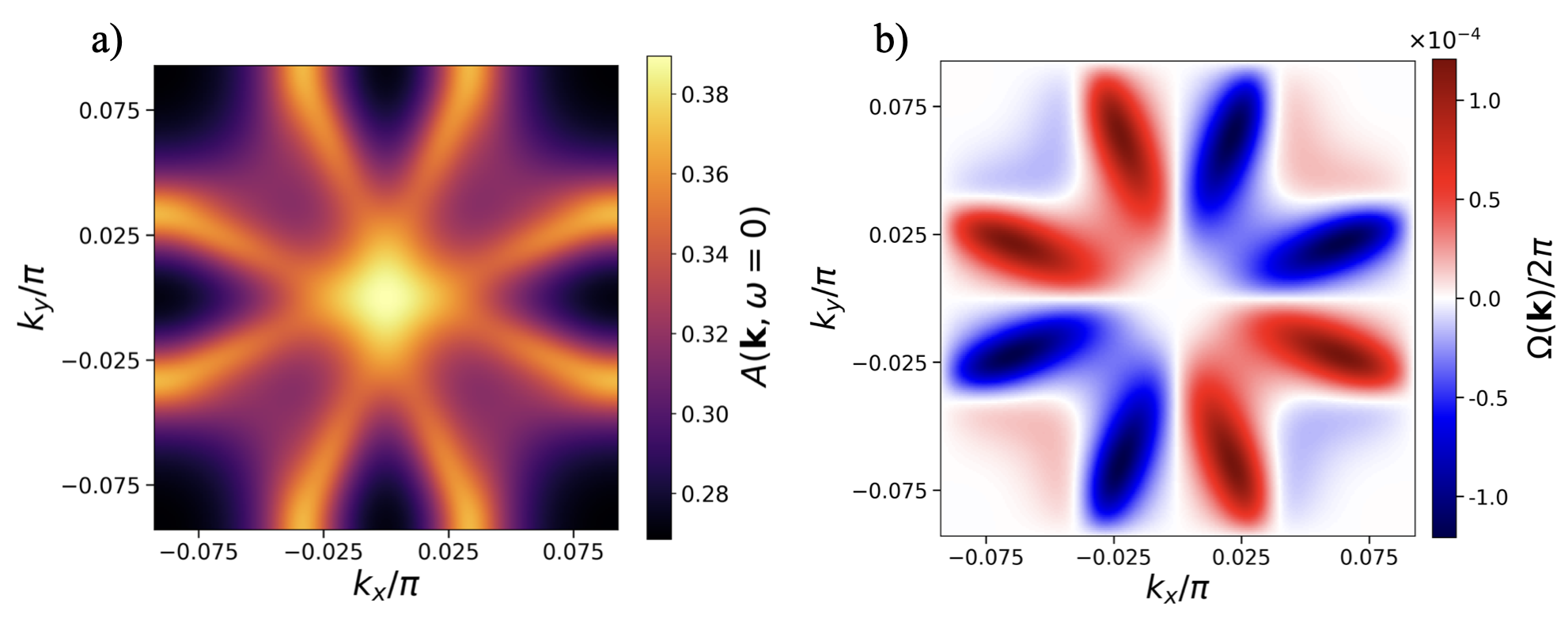}
\caption{(a) Spectral function at the Fermi energy, $A({\bf k},0)$, of the $s+i(d_1+d_2)$ state coexisting with AFM. The spectra is plotted in the first moire Brillouin zone of a twisted bilayer with 
$\theta=43.6^\circ$. As in Fig. \ref{fig:SM_berry} a) the spectra exhibits $C_4$ rotational invariance. (b) Berry curvature $\Omega(\mathbf{k})$ adding up to an overall zero Chern number $C=0$. We have fixed $t'=-0.5t$, $t_\perp=0.4$, $J=0.3t$, $\delta=0.15$ and broadening $\eta=10^{-4}$.
}
\label{fig:SM_berry_AFM}
\end{figure*}

The property behind this result, {\it i.e.} that in the $s+i(d_1+d_2)$ state a $\pi/2$ rotation acts as complex conjugation, is the same that makes the spectral function fourfold symmetric (Sec.~\ref{sec:supspectral}) even though the gap is only $C_2$ symmetric. The two quantities respond oppositely to complex conjugation: the Berry curvature, built from the imaginary part of overlaps of eigenstates, changes sign, whereas the spectral function is unchanged, since such a rotation leaves the diagonal elements of the Green's function invariant. Hence $C=0$, while $A_\sigma(C_4\mathbf{k},\omega)=A_\sigma(\mathbf{k},\omega)$ although the state itself breaks $C_4$ symmetry. Figure~\ref{fig:SM_berry} shows both aspects for $\Delta_s=\Delta_{d_1}=\Delta_{d_2}=0.5$ the curvature is concentrated in four hot spots at fourfold-symmetric positions on the Fermi surface, where $|\Delta_{\mathbf{k}}|$, and the spectral gap, is minimal, but their signs alternate under a $\pi/2$ rotation, so that their contributions to $C$ cancel.
 
At low doping the $s+i(d_1+d_2)$ state coexists with N\'eel order. Time reversal reverses every spin and thus turns the N\'eel pattern into its opposite, so $C_4\mathcal{T}$ is broken in the coexistence phase. In the absence of spin--orbit coupling, however, spins can be rotated independently of the lattice, and a $-\pi$ rotation about an axis perpendicular to the moments, $U_y(-\pi)$ for moments along $z$, flips them back: $\mathcal{T}U_y(-\pi)$ leaves the N\'eel order invariant, $i.e.$ $\mathcal{T}$ reverses $\mathbf{k}$, flips the spin and complex conjugates; $U_y(-\pi)$ undoes the spin flip. What remains is $\mathbf{k}\to-\mathbf{k}$ with complex conjugation and no spin flip. Thus, a $C_4$ rotation still acts on $\Delta({\bf k})$ as complex conjugation. The N\'eel order also doubles the unit cell, so each spin block becomes a $4\times4$ matrix in the basis $(c_{\mathbf{k}\uparrow},c_{\mathbf{k}+\mathbf{Q}\uparrow},c^\dagger_{-\mathbf{k}\downarrow},c^\dagger_{-\mathbf{k}-\mathbf{Q}\downarrow})$, with $\mathbf{Q}=(\pi,\pi)$ and $\mathbf{k}$ in the magnetic BZ (MBZ), $|k_x|+|k_y|\le\pi$. The staggered exchange field $h$ couples $\mathbf{k}$ and $\mathbf{k}+\mathbf{Q}$ through real matrix elements, and $C_4\mathbf{Q}$ equals $\mathbf{Q}$ up to a reciprocal lattice vector, so $h(C_4\mathbf{k})=h(\mathbf{k})^*$ still holds. The general proof then applies unchanged, with the integral over the MBZ and $\Omega$ the total curvature of the two occupied bands, $\Omega=\sum_{n\in{\rm occ}}\Omega_n=i\,\mathrm{Tr}\,P[\partial_{k_x}P,\partial_{k_y}P]$, where the projector $P$ onto the occupied states obeys $P(C_4\mathbf{k})=P(\mathbf{k})^*$. Hence the Chern number is forced to vanish, $C=0$, also in the coexistence phase.

In Fig. \ref{fig:SM_berry_AFM} we show results of $A({\bf k}, \omega)$ and the Berry curvature in the superconducting $s+i(d_1+d_2)$ state coexisting with AFM in  $\theta=43.6^\circ$ twisted bilayers.  $A({\bf k},0)$ displays $C_4$ symmetry as expected from our previous analysis based on the invariance of this coexisting state under a $C_4\mathcal{K}$ transformation. Also the Berry curvature displays an alternating sign structure under $\pi/2$ rotations leading to the overall $C=0$ Chern number.

\section{Edge States}

The unambiguous experimental characterization of topological superconductivity remains challenging. Unlike in Chern insulators, where the quantized Hall conductance directly determines the Chern number, the corresponding quantized response in two-dimensional chiral supercondcutors is the thermal Hall conductance, whose measurement remains experimentally demanding. Consequently, experimental searches largely rely on signatures of boundary or defect states at sample edges, interfaces and vortex cores.

\begin{figure}[h!]
\centering
\includegraphics[width=8.5cm]{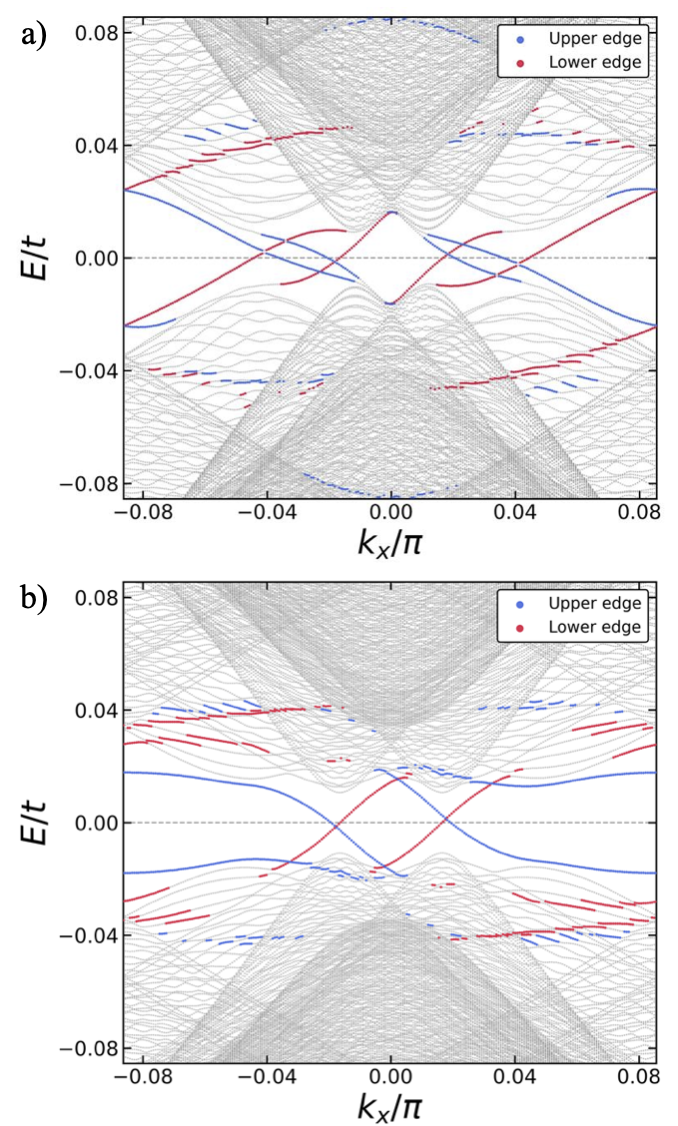}
\caption{Edge-state spectrum of a ribbon composed of $20$ moiré supercells along the open $y$ direction and subject to periodic boundary conditions along $x$, for $J=0.3t$, $t'=-0.5t$, and (a) $t_\perp=0.3t$, corresponding to the $d_1+id_2$ phase with Chern number $C=4$, and (b) $t_\perp=0.4t$, corresponding to the $s+d_1+id_2$ phase with $C=2$. Bulk states are shown in gray, whereas states localized at the lower and upper edges are shown in red and blue, respectively. In agreement with bulk-boundary correspondence, the $d_1+id_2$ phase in panel (a) exhibits twice as many chiral edge modes as the multicomponent phase in panel (b).}
\label{fig: edgestates_sup}
\end{figure}

We therefore, calculate the edge-state spectrum in a ribbon geometry that is periodic along $x$ and finite along the $y$-direction, diagonalizing the BdG Hamiltonian as a function of the conserved momentum $k_x$. Fig. \ref{fig: edgestates_sup} displays the ribbon edge-states for (a) $d_1+id_2$ and (b) $s+d_1+id_2$. The resulting in-gap states localized at opposite boundaries provide a direct test of the bulk-boundary correspondence expected from the bulk Chern number.

The coexistence of a gapped bulk spectrum and gapless chiral edge modes provides a spatially resolved signature of the topological phase
which can be probed by scanning tunneling microscopy (STM) through measurements of the local density of states (LDOS). While the LDOS is suppressed around the Fermi level in the bulk, finite in-gap spectral weight is expected at the edges and at interfaces separating regions with different topological invariants, owing to the presence of chiral boundary modes.

To make contact with such experiments, we have presented in Fig. 4 a) of the main text, the site-resolved
LDOS calculated for the same ribbon geometries employed in the edge-state analysis
and examined its spatial variation across the ribbon near the Fermi level.
Following the same procedure used to derive the spectral function, the
LDOS at site $i$ can be written as:
\begin{align*}
    \rho_i(\omega)
    =
    \frac{1}{N_{k_x}}
    \sum_{k_x,n,\sigma}
    \Big[
        &|u_{i\sigma n}(k_x)|^2
        \delta\!\left(\omega-E_n(k_x)\right)
        \\
        &+
        |v_{i\sigma n}(k_x)|^2
        \delta\!\left(\omega+E_n(k_x)\right)
    \Big],
\end{align*}
where the sum over $n$ runs over the positive-energy BdG
quasiparticle states.

\section{Critical Josephson current}

The expression for the Josephson current between the two layers can be obtained from charge conservation. Defining $\hat{N}_1=\sum_{i,\sigma}c^\dagger_{i1\sigma}c_{i1\sigma}$ as the particle-number operator of layer 1, the current density flowing from layer 1 to layer 2 reads:
\begin{equation}
    J(\varphi)=-\frac{e}{A}\frac{d \langle \hat{N}_1 \rangle}{dt}=\frac{ie}{\hbar A}\langle [\hat{N}_1, \mathcal{H}]\rangle,
\end{equation}
where $e>0$ is the elementary charge and $A=n^2+m^2$ is the area of the moiré unit cell in units of $a_0^2$. The phase variable, $\varphi=\varphi_2-\varphi_1$, from which the Josephson current is directly dependent, denotes the relative phase between the global superconducting order parameters of the two layers, $\Delta^{(l)}=|s_l+d_l|e^{i\varphi_l}$. Only the interlayer hopping contribution to $\mathcal{H}$ enters the commutator, yielding:
\begin{equation}
    J(\varphi)=\frac{2e}{\hbar A}\sum_{ij,\sigma} t_{ij}^\perp g_{ij,\sigma}^{(t)}\text{Im}\chi_{ij,\sigma}^{1\rightarrow 2}(\varphi),
\end{equation}
where $\chi_{ij,\sigma}^{1\rightarrow 2}=\langle c_{i1\sigma}^\dagger c_{j2\sigma}\rangle$. The critical Josephson current is then defined as the maximum magnitude of the current-phase relation:
\begin{equation}
    J_c=\max_{\varphi} |J(\varphi)|.
\end{equation}

To determine $J(\varphi)$ within our self-consistent mean-field approach, we first define a convenient gauge in which the phase bias can be explicitly controlled. Starting from the converged ground-state solution, we perform a layer-dependent gauge transformation and define the resulting state as the $\varphi=0$ reference configuration. This transformation reads:
\begin{align*}
    c_{i\sigma}^{(l)}\rightarrow c_{i\sigma}^{(l)}e^{i\varphi_l/2}.
\end{align*}

The overall phases of the intralayer global superconducting order parameters are thereby removed, $\Delta^{(l)}\rightarrow\Delta^{(l)}e^{-i\varphi_l}=|\Delta^{(l)}|$, while the interlayer pairing and hopping fields transfrom as $\Delta_{ij}^\perp\rightarrow\Delta_{ij}^\perp e^{-i(\varphi_1 + \varphi_2)/2}$, $t_{ij}^\perp\rightarrow t_{ij}^\perp e^{\pm i(\varphi_1-\varphi_2)/2}$, where the $+(-)$ sign in the later corresponds to hopping from layer 1 to layer 2 (layer 2 to layer 1). A finite Josephson bias is then introduced through an analogous gauge transformation of layer 1. For each value of $\varphi$, the intralayer pairing fields are kept real throughout the self-consistent evolution, preventing the imposed phase difference from relaxing while all the remaining mean-field parameters are allowed to relax self-consistently. The resulting converged solutions yield $J(\varphi)$ over the phase interval $[0,\pi)$, from which the critical Josephson current is obtained.

\bibliographystyle{apsrev4-1}
\bibliography{main}